\documentstyle[aps,eqsecnum,epsfig]{revtex}
\begin{document}

\draft

\author{B.~All\'es, M. Pepe}

\address{Dipartimento~di Fisica, Universit\`a di Milano-Bicocca \break
      and INFN Sezione di Milano, Milano, Italy}

\title{Four--loop free energy for the 2D $O(n)$ nonlinear
$\sigma$--model \break with 0--loop and 1--loop Symanzik 
improved actions\footnote{Preprint Bicocca--FT--99--04.}}
\maketitle

\vskip 1cm

\centerline{\it Abstract}
\begin{abstract}
We calculate up to four loops the free energy of the two--dimensional
(2D) $O(n)$ nonlinear 
$\sigma$--model regularized on the lattice with the 0--loop and 1--loop
Symanzik improved actions. An effective coupling constant based on this 
calculation is defined.
\end{abstract}

\pacs{05.50.+q; 11.15.Ha; 12.38.Bx; 75.10.Hk; 11.10.Kk}

\section{Introduction}

Numerical simulations on the lattice are a convenient method to
compute expectation values of dimensionful operators in a field
theory. To convert the expectation value obtained from the simulation
into a matrix element written in physical units we must fix the
lattice spacing $a$. In the present paper we will focus our attention
on the case of asymptotically free field theories.

The lattice scale $a$ is usually fixed through the integration of the
beta function of the theory. This function can be evaluated in
perturbation theory in which case $a$ is written also as a weak coupling
expansion (starting from three loops).
It is said that the asymptotic scaling regime has been attained if the
first universal terms in the perturbative expansion of the beta function
are enough to have a good knowledge of the lattice spacing $a$.
In principle we may expect that the better the weak expansion of 
the beta function is known, the more accurate the measurement of a
dimensionful physical observable will be. However this hope usually 
fails due to the bad convergence properties of the weak expansion. In
particular, the asymptotic scaling regime is barely achieved.

A cure to this problem may come from a redefinition of the expansion 
parameter in the perturbative series~\cite{gp,mpp,smm}. 
Usually the expansion parameter is the bare coupling constant $g$. 
However we may define any other parameter $g_E$
related to the old one by an expansion $g_E=g + O(g^2)$. If $g_E$ can
be numerically determined by some non--perturbative procedure then
the beta function, expressed as a power series in terms of $g_E$, 
can be regarded as a 
resummation where the non--perturbative effects have been
absorbed in the very definition of the  
new parameter $g_E$. The problem is the choice of such a 
non--perturbative procedure that would produce a good resummation and a 
rapid convergence of the power series in $g_E$. The calculation 
scheme where $g$ is substituted by $g_E$ shall hereafter be called 
``effective scheme''.

In Ref.~\cite{cp1} the 2D $O(n)$ nonlinear $\sigma$--model has been
analysed. This is an asymptotically free field theory,
which shares several physical properties with 
four--dimensional (4D) Yang--Mills theory, among others the spontaneous
generation of a mass in addition to the asymptotic freedom itself. 
In~\cite{cp1} the authors 
argued that the density of action (the internal 
energy) is a good operator to define $g_E$ (actually the choice of
this operator was proposed in the seminal papers~\cite{gp,mpp} for
4D gauge theories as well as for the 2D nonlinear $\sigma$--model). In
particular the corrections 
to asymptotic scaling in this effective scheme vanish in the large $n$ 
limit. In Ref.~\cite{abc}, also devoted to the study of the 2D nonlinear
$\sigma$--model, two different local operators were used to define the
new expansion 
parameter and in both cases a clear improvement was observed. Moreover the
results obtained for the measured quantities in the two
effective schemes were compatible with each other, which means that 
the implicit resummations of the series were equivalent. In the case
of gauge theories other types of redefinitions of the expansion
parameter can be used, see for instance~\cite{macklep}.

Nevertheless in many cases the lack of asymptotic scaling does not suffice
to explain the poor results obtained from a Monte Carlo
simulation. The lattice regularization of any operator (for example
the action) differs from its continuum counterpart by terms of higher
order in the lattice spacing $a$. If these terms are sizeable then the
scaling ratios of operator matrix elements as evaluated on the lattice 
do not behave as prescribed
by the continuum Renormalization Group equations, i.e.: they do not
show physical scaling. This is another source of systematic errors in
the Monte Carlo determination of any dimensionful observable.
This problem can be mitigated if the Monte Carlo simulation is
performed with a Symanzik improved action. 

This difficulty occurs again in the 2D nonlinear
$\sigma$--model. For instance, the calculation of the mass gap in
this model with $O(3)$ symmetry 
has yielded results~\cite{st,fgmo,fo,hm1,hm2,fmptp,uw0,hn,abf} 
in conflict with the exact analytical value $m_{HMN}$ of Ref.~\cite{hmn}.
In Ref.~\cite{uw1} an energy--based effective scheme was used to
extract the mass gap but the deviation between the Monte Carlo and
exact results was still beyond 10\%. 
A definite improvement was obtained only after the problem 
concerning the lack of physical scaling was also tackled. In
Ref.~\cite{abc} a combination of an effective scheme together with a
Symanzik improved action allowed an agreement between the Monte Carlo
and the exact results well within 2--3\% 

Therefore one of the main conclusions of Ref.~\cite{abc} 
was that the combination
of an effective scheme together with a Symanzik improved action may allow to
obtain results reasonably clean from all lattice artifacts (assuming that
the lattice volume is large enough). 
Our purpose is to further check this conclusion. To this end we want
to calculate the mass gap of the 2D $O(3)$ nonlinear 
$\sigma$--model on the lattice by simulating two improved actions, 
the tree--order Symanzik and the 1--loop Symanzik actions~\cite{ks1,ks2}. 
Let us call $\Delta m$ the difference between the mass extracted from
a Monte Carlo simulation and the exact result of Ref.~\cite{hmn}. 
We want to study the dependence of $\Delta m$ on the level of improvement
in the action as well as on the number of corrective terms to the
asymptotic scaling in the perturbative expansion of the lattice
spacing. Firstly we 
need to calculate these corrections in perturbation theory
and this is done in Ref.~\cite{ap} where we calculate the lattice beta 
function $\beta^L$ for the two above--mentioned Symanzik 
actions up to four loops in terms of the bare coupling $g$. In the
present paper we will show the result for $\beta^L$ in 
terms of an effective coupling $g_E$ defined through
the internal energy $E$. We will perform the calculation for a generic
symmetry group $O(n)$, the specialization to the case $n=3$ being
trivial. 

Our interest is actually twofold 
because a high precision determination of the mass
in the model would also contribute to settle a long--standing debate
about the validity of the $m_{HMN}$ value~\cite{ps}.

In the following section we give the expressions of the
two Symanzik improved actions and the explicit form of the operators
for $E$. We also briefly describe our calculation procedure.
In the third section we introduce some notation and a list of useful
identities that we have used. The final results for the internal energy $E$
will be given in section 4 together with the calculation of the lattice 
beta function in terms of $g_E$. A summary of the results and
a brief account of the checks performed on them is given in the conclusions.
Several appendices are devoted to the finite integrals that appear in
the final expressions; we list these integrals and their numerical values, 
explaining how they were evaluated and showing some identities
which relate them.

\section{The calculation}

In this section we give an outline of the calculation method that we have
followed. The action of the nonlinear $O(n)$--symmetric $\sigma$--model 
on the lattice can
be written in many different ways. All of them share the same na\"{\i}ve 
continuum limit, i.e. the action of the model in the continuum
\begin{equation}
 S^{\rm continuum} = {1 \over 2 g} \int \hbox{d}^2 x \sum_\mu
       \left(\partial_\mu \, \vec{\phi}(x) \right)^2 \;,
\label{scontinuum}
\end{equation}
where $g$ is the coupling constant and $\vec\phi$ is a scalar field with
$n$ components constrained by the condition 
\begin{equation}
 \left(\vec\phi(x)\right)^2 = 1 \qquad\qquad \hbox{for all $x$}\; .
\label{constraint}
\end{equation}
In this work we shall consider three actions on 
the lattice. The standard action is the simplest one but yields the
poorest results in simulations; on the other hand
the Symanzik improved actions allow
a progressive elimination of lattice artifacts as powers of the lattice
spacing $a$ and its logarithm $\log a$~\cite{ks1,ks2}. These actions are
\begin{eqnarray}
 S^{\rm standard} &=&  {a^2 \over g} \sum_{x} {1 \over 2}
                       \vec\phi(x) \cdot K_1\cdot  \vec\phi(x)
                       \;,\nonumber \\
 S^{\rm 0-Symanzik} &=& {a^2 \over g} \sum_{x} 
                     \left( {2 \over 3} \vec\phi(x) \cdot K_1\cdot  \vec\phi(x) -
                            {1 \over 24} \vec\phi(x) \cdot K_2\cdot  \vec\phi(x)
                     \right) \; ,\nonumber \\
 S^{\rm 1-Symanzik} &=&   {a^2 \over g} \sum_x \bigg[ \;{1 \over 2}
         \vec\phi(x) \cdot K_1\cdot  \vec\phi(x) -
         a^2c_5 \;g\left(K_1\cdot  \vec\phi(x)\right)^2 -
         a^2\left(c_6 \; g\,- {1 \over 24}\right) 
        \sum_\mu \left( \partial^+_\mu\partial^-_\mu \vec\phi(x)\right)^2 
      \nonumber \\
 && \qquad  - a^2c_7 \;g\left(\vec\phi(x) \cdot K_1\cdot  
         \vec\phi(x)\right)^2 - a^2c_8\;g
         \sum_\mu \left(\vec\phi(x) \cdot
         \partial^+_\mu\partial^-_\mu \vec\phi(x)\right)^2 
 \nonumber \\
 && \qquad      -  {1 \over 16} a^2 c_9 \;g\sum_{\mu\nu} \left(
        \left(\partial^+_\mu + \partial^-_\mu\right) \vec\phi(x) \cdot
        \left(\partial^+_\nu + \partial^-_\nu\right) \vec\phi(x) \right)^2 \bigg]
      \; ,
\label{slattice}
\end{eqnarray}
where the superscript $i$--Symanzik denotes the 
$i$--loop Symanzik improved action. The several operators in this
equation are defined by
\begin{eqnarray}
 K_1\cdot \vec\phi(x) & \equiv & {1\over a^2}\sum_\mu \left( 2 \vec\phi(x) - 
               \vec\phi(x+\widehat\mu) - 
               \vec\phi(x-\widehat\mu) \right)\; ,\nonumber \\
 K_2\cdot \vec\phi(x) & \equiv & {1\over a^2}\sum_\mu \left( 2 \vec\phi(x) - 
               \vec\phi(x+2\,\widehat\mu) - 
               \vec\phi(x-2\,\widehat\mu) \right)\; ,\nonumber \\
 \partial^+_\mu \vec\phi(x) &\equiv& {1\over a}\left(\vec\phi(x+\widehat\mu) - 
             \vec\phi(x)\right) \; ,\nonumber \\
 \partial^-_\mu \vec\phi(x) &\equiv& {1\over a}\left(\vec\phi(x) - 
             \vec\phi(x-\widehat\mu) \right)\; .
\label{operators}
\end{eqnarray}
The set of coefficients $\{c_i\}_{i=5,...,9}$ is determined by the
Symanzik improvement program at one loop~\cite{ks2,bmms} (there is a
discrepancy between the numerical values of $c_6$ reported 
in~\cite{bmms} and~\cite{ks2}, we agree with the
latter, see~\cite{montanari}). We shall often 
refer to these actions as standard action, 0--loop action and 1--loop
action respectively.

The 1--loop action can be written in a more convenient form (mainly for
the Monte Carlo simulation) after 
introducing the operators of Eq.(\ref{operators}) into Eq.(\ref{slattice}),
\begin{eqnarray}
 S^{\rm 1-Symanzik} &=& - \sum_x \Biggl\{{1 \over g}
           \sum_\mu \left( {4 \over 3} \vec\phi(x) \cdot \vec\phi(x+\widehat\mu) -
                   {1 \over 12} \vec\phi(x) \cdot \vec\phi(x+2\widehat\mu)\right) +
             \nonumber \\
    &&       c_5 \Bigg[2 \sum_\mu \vec\phi(x) \cdot \vec\phi(x+2\widehat\mu) -
                     16 \sum_\mu \vec\phi(x) \cdot \vec\phi(x+\widehat\mu) +
              \nonumber \\
     && \qquad     2 \sum_{\mu\not=\nu} \vec\phi(x+\widehat\mu) \cdot
                       \vec\phi(x+\widehat\nu) + 2 \sum_{\mu\not=\nu}
                      \vec\phi(x) \cdot \vec\phi(x+\widehat\mu+\widehat\nu)\Bigg] +
          \nonumber \\
    && c_6 \Bigg[ 2 \sum_\mu \vec\phi(x) \cdot \vec\phi(x+ 2\widehat\mu) -
              8  \sum_\mu \vec\phi(x) \cdot \vec\phi(x+\widehat\mu) \Bigg] +
          \nonumber \\
    && c_7 \Bigg[ -16 \sum_\mu \vec\phi(x) \cdot \vec\phi(x+\widehat\mu) +
           2 \sum_\mu \left( \vec\phi(x) \cdot \vec\phi(x+\widehat\mu)\right)^2 +
       \nonumber \\
  &&\qquad 2 \sum_\mu \left( \vec\phi(x) \cdot \vec\phi(x+\widehat\mu)\right) \,
               \left( \vec\phi(x+\widehat\mu) \cdot 
                      \vec\phi(x+2\widehat\mu)\right) +
          \nonumber \\
    && \qquad \sum_{\mu\not=\nu} \Big(
           \left( \vec\phi(x) \cdot \vec\phi(x+\widehat\mu)\right) \,
           \left( \vec\phi(x) \cdot \vec\phi(x+\widehat\nu)\right) +
           \left( \vec\phi(x) \cdot \vec\phi(x+\widehat\mu)\right) \,
           \left( \vec\phi(x) \cdot \vec\phi(x-\widehat\nu)\right) +
          \nonumber \\
 &&  \qquad\qquad \left( \vec\phi(x) \cdot \vec\phi(x-\widehat\mu)\right) \,
           \left( \vec\phi(x) \cdot \vec\phi(x+\widehat\nu)\right) +
           \left( \vec\phi(x) \cdot \vec\phi(x-\widehat\mu)\right) \,
           \left( \vec\phi(x) \cdot \vec\phi(x-\widehat\nu)\right) \Big) \Bigg] +
          \nonumber \\
    && c_8 \Bigg[ 2 \sum_\mu \left(\vec\phi(x)\cdot
                                   \vec\phi(x+\widehat\mu)\right)^2 -
             8 \sum_\mu \vec\phi(x) \cdot \vec\phi(x+\widehat\mu) + 
             2 \sum_\mu \left(\vec\phi(x) \cdot \vec\phi(x+\widehat\mu) \right) \,
                        \left(\vec\phi(x) \cdot 
                              \vec\phi(x-\widehat\mu) \right)\Bigg]+
          \nonumber \\
    && {c_9 \over 16}  \Bigg[ 4 \sum_\mu 
        \left(\vec\phi(x+\widehat\mu) \cdot \vec\phi(x-\widehat\mu)\right)^2 - 
  8 \sum_\mu \vec\phi(x) \cdot \vec\phi(x+2\widehat\mu) + \nonumber \\
    && \qquad \sum_{\mu\not=\nu} \bigg(
     \left( \vec\phi(x+\widehat\mu) \cdot \vec\phi(x+\widehat\nu) \right)^2 +
     \left( \vec\phi(x+\widehat\mu) \cdot \vec\phi(x-\widehat\nu) \right)^2 +
     \nonumber \\
  &&\qquad\qquad \left( \vec\phi(x-\widehat\mu) \cdot 
                        \vec\phi(x+\widehat\nu) \right)^2 +
     \left( \vec\phi(x-\widehat\mu) \cdot 
            \vec\phi(x-\widehat\nu) \right)^2 \bigg) +
          \nonumber \\
    && \qquad 2 \sum_{\mu\not=\nu} \bigg(
      \left(\vec\phi(x+\widehat\mu) \cdot \vec\phi(x+\widehat\nu)\right) \,
      \left(\vec\phi(x-\widehat\mu) \cdot \vec\phi(x-\widehat\nu)\right) +
          \nonumber \\
    &&\qquad\qquad \left(\vec\phi(x+\widehat\mu) \cdot 
                         \vec\phi(x-\widehat\nu)\right) \,
      \left(\vec\phi(x-\widehat\mu) \cdot \vec\phi(x+\widehat\nu)\right)  -
          \nonumber \\
    && \qquad \qquad
      \left(\vec\phi(x+\widehat\mu) \cdot \vec\phi(x+\widehat\nu)\right) \,
      \left(\vec\phi(x+\widehat\mu) \cdot \vec\phi(x-\widehat\nu)\right) -
          \nonumber \\
    &&\qquad\qquad \left(\vec\phi(x+\widehat\mu) \cdot 
                         \vec\phi(x+\widehat\nu)\right) \,
      \left(\vec\phi(x-\widehat\mu) \cdot \vec\phi(x+\widehat\nu)\right) -
          \nonumber \\
    &&\qquad\qquad \left(\vec\phi(x-\widehat\mu) \cdot 
                         \vec\phi(x-\widehat\nu)\right) \,
      \left(\vec\phi(x+\widehat\mu) \cdot \vec\phi(x-\widehat\nu)\right) -
          \nonumber \\
    &&\qquad\qquad  \left(\vec\phi(x-\widehat\mu) \cdot 
                          \vec\phi(x-\widehat\nu)\right) \,
      \left(\vec\phi(x-\widehat\mu) \cdot 
            \vec\phi(x+\widehat\nu)\right) \bigg) \Bigg]
     \Biggr\}\; .
\label{s1loop}
\end{eqnarray}

We notice that this expression differs from the one shown in Table~2
of Ref.~\cite{bmm}. We can only say that we have carefully checked our
Eq.(\ref{s1loop}) and that it coincides with the action given for
example in Ref.~\cite{tk}.

The terms proportional to the 
coefficients $c_i$ are of order $O(g)$. Therefore 
the 1--loop action is equal to 
the 0--loop action plus a sum of terms of higher order in $g$. This fact
allows us to compute any perturbative quantity at $k$ loops for the 1--loop 
action as the sum of the analogous quantity for the 0--loop action plus a set of 
diagrams with at most $\left(k-1\right)$ loops.

To the actions showed in Eqs.(\ref{slattice}) and 
(\ref{s1loop}) we must still add
another term. The constraint shown in Eq.(\ref{constraint}) is introduced 
under the path integral representation of the theory as a Dirac delta.
Hence the partition function $Z$ is written as
\begin{equation}
Z\equiv \int {\cal D}\vec\phi(x)\; 
   \prod_x \delta\left(\left(\vec\phi(x)\right)^2-1\right)\; 
   \hbox{e}^{-S}\;,
\end{equation}
where $S$ indicates any of the lattice actions that we consider in the present
paper. This constraint can be solved by rewriting the $n$--component
field $\vec{\phi}$ in terms of a new $\left(n-1\right)$--component 
field $\vec\pi$ in the following way
\begin{eqnarray}
 \vec\phi(x) &=& \Big(\phi_1(x),\;...,\; \phi_n(x)\Big) \;\;
      \longrightarrow \left(\pi_1(x),\;...,\; \pi_{n-1}(x),\;
      \sqrt{1 - \pi_1(x)^2 - \;...\; \pi_{n-1}(x)^2}\;\right)
     \nonumber \\
     &\equiv& \left(\vec\pi(x),\; \sqrt{1 - \vec\pi(x)^2}\right) \; .
\label{phipi}
\end{eqnarray}
This transformation introduces a change of variables 
in the functional integration whose
jacobian becomes a new term to be added to the original action as a
measure action~\cite{bzjg},
\begin{equation}
 S  \longrightarrow S + \sum_x \log \sqrt{1 - \vec\pi(x)^2} \; .
\label{smeasure}
\end{equation}
Perturbation theory is developed for the $(n-1)$--component field
$\vec\pi(x)$ around the trivial configuration $\vec\pi(x)=0$ for all $x$.
The contribution to perturbative expansions from 
the measure term vanishes in the 
continuum when the divergences are dimensionally regularized~\cite{thooft}.

In this paper we will compute the average density of action (or internal
energy) $E$, up to fourth order in perturbation theory 
for all three actions on the lattice. We will begin by computing the
free energy $F$ to four loops, from which the internal energy reads
\begin{equation}
 E  = w_0 + {1\over 2}{\partial \over \partial \left(1/g\right)} F\;,
 \qquad \qquad \qquad F\equiv \lim_{L\rightarrow\infty}
 {1\over L^2}\log Z \;,
\label{derivative}
\end{equation}
where $L^2$ is the volume of a two--dimensional square lattice of side
length $L$ and $Z$ is the partition function.
The constant $w_0$ is determined in such a way that the operators
representing $E$ are (there is no summation in the index $\mu$)
\begin{eqnarray}
 E^{\rm standard}&=& \langle \vec\phi(0) \cdot \vec\phi(0+\widehat{\mu})\rangle 
     \;, \nonumber \\
 E^{\rm 0-Symanzik}&=& \langle {4\over 3}\vec\phi(0) 
            \cdot \vec\phi(0+\widehat{\mu})
            -{1\over 12} \vec\phi(0) \cdot \vec\phi(0+2\widehat{\mu})\rangle 
     \;, \nonumber \\
 E^{\rm 1-Symanzik}&=&\langle {4\over 3}\vec\phi(0) 
            \cdot \vec\phi(0+\widehat{\mu})
            -{1\over 12} \vec\phi(0) \cdot
     \vec\phi(0+2\widehat{\mu})\rangle \;.
\label{representing}
\end{eqnarray}
Then $w_0^{\rm standard}=1$ for the standard action and
$w_0^{\rm 0-Symanzik}=w_0^{\rm 1-Symanzik}=15/12$ for the two Symanzik
actions.

In perturbation theory $F$ is computed through the sum of the
connected Feynman diagrams shown in Figs.~1--6. The analytic
expressions of these diagrams are plagued with infrared
(IR)--divergent integrals. 
An usual method to deal with these IR divergences is the introduction
of an external source $h$ coupled to the scalar field 
$\vec\phi$~\cite{bzjg,david}. We have chosen a different IR regulator:
we have put the model into a square box of size $L$ with periodic
conditions on the boundaries. Then we have applied
finite--size perturbation theory following Ref.~\cite{hasen2}.
This procedure has two consequences: firstly all zero modes can be
excluded from the sums over momenta and secondly the action contains a
new term which comes from a Faddeev--Popov determinant,
\begin{equation}
 S_{FP} = - \left(n-1\right) \log \left( 
            \sum_x \sqrt{1 - \vec{\pi}(x)^2} \right)\;.
\label{faddeevv}
\end{equation}
We have explicitly checked that the diagrams containing 
vertices from $S_{FP}$ give rise to contributions which 
vanish in the thermodinamic limit $L \rightarrow \infty$ order by
order up to four loops (individual diagrams give rise to
non--vanishing terms but they cancel when adding up all diagrams 
at each order). We have not
included this list of diagrams in Figs.~1--6 because in this paper
we give the results for infinite--size lattices (in any case their
contribution is rather small: for $L=100$ it amounts to $O(10^{-5})\;$).

The diagrams in Figs.~1--6 lead to finite sums over internal
momenta. Some of the sums become
IR--divergent integrals in the limit $L\rightarrow\infty$.
We have worked out these sums in order to separate their finite
contribution from the IR--divergent one. In order to
isolate these divergences we have performed only algebraic
manipulations. We have avoided other kinds of manipulations, like
applying derivatives to the integrands. Then the divergent pieces
exactly cancel among themselves order by order. At the end, only the
convergent sums remain (sums which in the limit $L \rightarrow \infty$
turn into IR--finite integrals) and we can safely remove the IR
regulator by sending $L$ to infinity. These finite integrals are then
calculated as explained in Appendix~C. 

The calculation up to four loops for the standard action is known in the
literature~\cite{bl,cp1,ml,abc}. We have checked it obtaining
analytical results in agreement.
The calculation for the 0--loop Symanzik action is known only up to
third order~\cite{dgfpv,abc} and we have also successfully checked it.
Therefore the new results of the present paper are the 
fourth order for the 0--loop Symanzik action and the full calculation to all
orders, up to four loops, for the 1--loop improved Symanzik action. 
For completeness, in section 4 we will give the final results for all
actions.

\section{Notation and Identities}

In this section we will introduce some standard notation for the lattice
perturbative calculations. Unless otherwise stated, we set the lattice
spacing $a=1$. The sine of half of the $\mu$ component
of a momentum $p$ is denoted by
\begin{equation}
 \widehat{p}_\mu \equiv 2 \, \sin {p_\mu \over 2} \; .
\end{equation}
The inverse propagator in the standard action is 
$\widehat{p}^2 \equiv \sum_\mu \widehat{p}_\mu^2$. For the 
Symanzik action we introduce the notation
\begin{equation}
 \Box_p\equiv \sum_\mu \widehat{p}_\mu^4\; ,
\end{equation}
and then the inverse Symanzik propagator, denoted by $\Pi_p$, reads
\begin{eqnarray}
 \Pi_p &\equiv& \widehat{p}^2 + {1\over 12} \Box_p \equiv
  \sum_\mu \Pi_p^\mu \;,\nonumber \\
 \Pi_p^\mu &\equiv& \widehat{p}^2_\mu + {1\over 12} \widehat{p}_\mu^4 \;.
\label{propagatorsym}
\end{eqnarray}
This propagator is the same for the two Symanzik actions.

Several identities among momenta are helpful to separate the finite 
and divergent contributions to $E$. One of these
identities involve standard propagators~\cite{cp1},
\begin{equation}
 \widehat{(p+q)}^2 + \widehat{(p+k)}^2 + \widehat{(p+r)}^2 = 
 \widehat{p}^2 +\widehat{q}^2 +\widehat{k}^2 +\widehat{r}^2 -
 \Sigma_{pqkr}\;, \qquad \qquad
 \Sigma_{pqkr}\equiv\sum_\mu \widehat{p}_\mu \widehat{q}_\mu 
 \widehat{k}_\mu \widehat{r}_\mu \;,
\end{equation}
and it is valid if $p+q+k+r=0$. For the calculation using the Symanzik
actions other relationships are needed, for instance~\cite{abc},
\begin{eqnarray}
 \Pi_{p+q} +\Pi_{p+k} +\Pi_{p+r} &=&
 \Pi_p +\Pi_q +\Pi_k +\Pi_r - \Sigma^S_{pqkr}\;,
 \nonumber \\
 \Sigma^S_{pqkr}&\equiv& {4\over 3} \left( \sum_\mu 
 \widehat{p}_\mu \widehat{q}_\mu \widehat{k}_\mu \widehat{r}_\mu -
 \sum_\mu 
 \sin p_\mu \; \sin q_\mu \; \sin k_\mu \; \sin r_\mu \right)\; ,
\label{identities3}
\end{eqnarray}
which again requires that $p+q+k+r=0$. Moreover, in the calculation
of tadpole diagrams we have used
\begin{eqnarray}
 \widehat{p+q}^2 &=& \widehat{p}^2 + \widehat{q}^2 - {1\over 4} 
                   \widehat{p}^2 \widehat{q}^2 + \hbox{odd terms,}\nonumber \\
 \left( \widehat{p+q}^2 \right)^2 &=&
 \left(\widehat{p}^2\right)^2 + \left(\widehat{q}^2\right)^2 + {1 \over 8} 
 \left[\left(\widehat{p}^2\right)^2 \left(\widehat{q}^2\right)^2 -
      \left( \left(\widehat{p}^2\right)^2 - \Box_p\right) \Box_q -
      \left( \left(\widehat{q}^2\right)^2 - \Box_q\right) \Box_p\right] 
       \nonumber \\
 && \qquad + 2 \left( \widehat{p}^2 - {1\over 4} \Box_p \right)
             \left( \widehat{q}^2 - {1\over 4} \Box_q \right) +
           2 \widehat{p}^2 \widehat{q}^2 - {1\over 2} 
             \left(\widehat{p}^2\right)^2 \widehat{q}^2 -{1\over 2} 
             \left(\widehat{q}^2\right)^2 \widehat{p}^2 + \hbox{odd terms,}
       \nonumber \\
 \Pi_{p+q} &=& \Pi_{p} + \Pi_{q} - {1\over 12} \Pi_p \Box_q 
     - {1\over 12} \Pi_q \Box_p + {5\over 144} \Box_p \Box_q + \hbox{odd terms,}
       \nonumber \\
 \Box_{p+q} &=& \Box_p + \Box_q - {5 \over 4} \Pi_p \Box_q  
         - {5 \over 4} \Pi_q \Box_p + 3 \Pi_p \Pi_q + {7 \over 16}
         \Box_p \Box_q + \hbox{odd terms.}
\label{identities2}
\end{eqnarray}

The Feynman diagrams necessary for the calculation are shown in Figures~1--6.

Let us show two partial calculations as examples of the procedure we have
followed. On finite lattices of side length $L$ any component of a
momentum $p$ can take $L$ discrete values, for instance the first
component $p_1=2\pi\ell_1/L$ ($\ell_1=0$, 1, ..., $L-1$). Therefore
the sums over momenta are
\begin{equation} 
 {1\over L}\sum_{\ell_1=0}^{L-1} \negthinspace\negthinspace{}^{*} \;\;
 {1\over L}\sum_{\ell_2=0}^{L-1} \negthinspace\negthinspace{}^{*} \; ,
\label{finitesum}
\end{equation}
and become integrals in the limit $L\rightarrow\infty$,
\begin{equation}
 \int_{-\pi}^{+\pi}  {{{\rm d^2} p} \over  \left(2 \pi\right)^2} \; .
\end{equation}
In Eq.(\ref{finitesum}) the zero mode $\ell_1=\ell_2=0$ must be
excluded as prescribed in Ref.~\cite{hasen2}. This is the meaning of
the stars in Eq.(\ref{finitesum}). In the following we use the
shorthand 
\begin{equation}
  {\sum}_p \; ,
\label{sumshort}
\end{equation}
to denote the sum in Eq.(\ref{finitesum}) when it sums over the
momentum $p$. Besides, although at finite $L$ we are dealing with
discrete sums, we will often call them ``integrals'' and the
expression summed will often be named ``integrand''.

Momentum conservation is expressed through a Kronecker delta,
$L^2\,\delta^2(p+q+k+\cdots)$. $p$, $q$, $k$, ... are momenta which
satisfy $p_1=2\pi\ell_1/L$, etc. The argument of the delta function is
assumed to be periodic modulus $2\pi$. The sum in Eq.(\ref{sumshort}),
or in Eq.(\ref{finitesum}), acting on a delta function leads to
\begin{equation}
 {\sum}_p \; L^2\, \delta^2(p+q+k+\cdots) =
 1 - {1\over L^2} \;L^2 \, \delta^2(q+k+\cdots) \;.
\label{sumoverdelta}
\end{equation}
The additional $O(1/L^2)$ contribution in the r.h.s. of this equation can
produce terms like
\begin{equation}
 {1\over L^2}\; {\sum}_p\; {1\over\left(\Pi_p\right)^2}\;,
 \qquad\qquad\qquad\qquad
 {1\over L^2}\; {\sum}_p {\sum}_q {\sum}_k\; {1\over \Pi_p\;\Pi_q\;\Pi_k}
 \; L^2\, \delta^2(p+q+k) \; ,
\label{examplesexclusive}
\end{equation}
which are finite after the removal of the IR--regulator. These terms
are exclusive of the finite size $L$ regularization: notice for
example that they cannot be
expressed as usual integrals, neither after the $L\rightarrow\infty$
limit. We have checked that such terms cancel out when we sum up the
contributions from all diagrams at each order.

Terms like those in Eq.(\ref{examplesexclusive}), and similar ones
coming from the Faddeev--Popov action (\ref{faddeevv}), yield finite
contributions to the final result in the 1D $O(n)$ model~\cite{hasen2}.
In the 2D model such terms cancel out at least up to fourth order. 

To the diagram 6 of Fig~\ref{fig:6} it contributes the
coefficient $c_9$ of the 1--loop improved action through the vertex
\begin{eqnarray}
   {1\over 2} \; c_9 \;  & &
{\sum}_p\;
{\sum}_q\;
{\sum}_k\;
{\sum}_r\;
{\sum}_s\;
{\sum}_t\;
 L^2\; \delta^2(p+q+k+r+s+t) \nonumber \\
 && \qquad \qquad \qquad \qquad \times \; \sum_{\mu\nu} \sum_{abc}  
     \pi^a(p) \, \pi^a(q) \, \pi^b(k) \, 
     \pi^b(r) \, \pi^c(s) \, \pi^c(t) \, 
     \sin p_\mu \; \sin q_\nu\;
     \sin\left( k+r\right)_\mu \; \sin\left( s+t\right)_\nu \;,
\end{eqnarray}
where $p$, $q$, $k$, $r$, $s$ and $t$ are the momenta which satisfy 
$p_1=2\pi\ell_1/L$, etc. and $a$, $b$ and
$c$ the $O(n)$ indices (running from 1 to $n-1$). 
To simplify the notation we use the same symbol
to indicate the field in coordinate space $\vec\pi(x)$ and in momentum
space $\vec\pi(p)$. The contraction of the
six legs to produce three tadpoles can be done in three topologically
non equivalent ways and leads to
\begin{eqnarray}
  c_9\  g^3 \; \sum_{\mu\nu}\; \Bigg\{  &&
  \left(n-1\right)^2 \;
{\sum}_p\;
{\sum}_q\;
{\sum}_k\;
  {{\sin p_\mu \; \sin p_\nu \; \sin \left( q+k\right)_\mu 
    \sin \left(q+k\right)_\nu} \over \Pi_p \Pi_q \Pi_k }  
  \nonumber \\
  +\;& 2 &\; \left(n-1\right) \;
{\sum}_p\;
{\sum}_q\;
{\sum}_k\;
  {{\sin p_\mu \; \sin q_\nu \; \sin \left( q+k\right)_\mu 
    \sin \left(p+k\right)_\nu} \over \Pi_p \Pi_q \Pi_k } 
  \nonumber \\
  +\;& 2 &\; \left(n-1\right) \;
{\sum}_p\;
{\sum}_q\;
{\sum}_k\;
  {{\sin p_\mu \; \sin q_\nu \; \sin \left( p+k\right)_\mu 
    \sin \left(q+k\right)_\nu} \over \Pi_p \Pi_q \Pi_k } \Bigg\} \;.
\label{sunday30}
\end{eqnarray}
The first integral is non zero only if $\mu=\nu$ and, by using
$\sin^2 p_\mu =  \widehat{p}^2_\mu - 1/4\; \widehat{p}^4_\mu$ and
the first and last identities in Eq.(\ref{identities2}), gives
\begin{equation}
c_9\;g^3\;{1\over 2}\;\left(n-1\right)^2\; \left( 1- {1\over 3} Y_1\right)^2
   \left( 2 Z_1 - 1 + {1\over 3}  Y_1\right) + 
   O\left({\log L\over L^2}\right)\; ,
\end{equation}
where $Y_i$ in the limit $L\rightarrow\infty$ 
are finite integrals defined and calculated in Appendix~A
and $Z_i$ are the above--mentioned sums which in the thermodynamic
limit diverge and that at the end of the calculation must disappear,
\begin{equation}
 Z_i \equiv {\sum}_p
 \left({1\over \Pi_p}\right)^i \; .
\end{equation}
Notice that this expression is well defined as long as $L$ is finite
because the zero mode $\Pi_p=0$ is missing. It becomes an ill--defined
integral only in the thermodynamic limit. For large $L$ we have that
$Z_1\sim\log L$, $Z_2\sim L^2$, $Z_3\sim L^4$, etc.

The second integration in Eq.(\ref{sunday30}) yields, after some
algebra and taking into account again that it vanishes when $\mu\not=\nu$, 
\begin{eqnarray}
 2\;c_9\;g^3 \; \left(n-1\right) \;
{\sum}_p\;
{\sum}_q\;
{\sum}_k &&
  {{\sum_\mu \sin^2 p_\mu \; \sin^2 q_\mu \; \sin^2 k_\mu} \over 
    \Pi_p\; \Pi_q\; \Pi_k }
  \nonumber \\
 & =& c_9\;g^3\;{1\over 2} \; \left(n-1\right) \;
    \left(1 - {1\over 3} Y_1\right)^2 \left( 2 Z_1 - 1 + {1\over 3}
  Y_1\right) +
  O\left({\log L\over L^2}\right)\; .
\end{eqnarray}
Finally the third integral in Eq.(\ref{sunday30}) needs some algebra to
eliminate several odd terms like $\sum_p \sin p_\mu / \Pi_p$. After
this work, it can be rewritten as 
\begin{eqnarray}
 2\;c_9\;g^3\; \left(n-1\right) \;
{\sum}_p\;
{\sum}_q &&
{\sum}_k \;
  {{\sum_{\mu\nu} \sin^2 p_\mu \; \sin^2 q_\nu \; \cos k_\mu \cos
  k_\nu} \over \Pi_p\; \Pi_q\; \Pi_k }
  \nonumber \\
 & =& c_9\;g^3\; {1\over 2} \; \left(n-1\right) \;
    \left(1 - {1\over 3} Y_1\right)^2 
    \left( 4 Z_1 - {5\over 4} + {1\over 6} Y_1 + {1\over 576} 
    Y_{2,1}\right) +
    O\left({\log L\over L^2}\right)\; ,
\end{eqnarray}
where the property $\cos k_\mu = 1 - 1/2\; \widehat{k}^2_\mu$ was used.

The second example of calculation is taken from the coefficient of
$c_9$ in the diagram 5 of Fig.~\ref{fig:6}. The vertices with four
legs are
\begin{eqnarray}
   {1\over 16} \; c_9 \;  & &
{\sum}_p\;
{\sum}_q\; 
{\sum}_k\; 
{\sum}_r\; 
 L^2 \;\delta^2(p+q+k+r) \nonumber \\
 && \qquad \qquad \times \;  \sum_{ab}  
     \pi^a(p) \, \pi^a(q) \, \pi^b(k) \, \pi^b(r) \, 
     \left[ \left(\widehat{p+k}^2\right)^2 +
            \left(\widehat{p-k}^2\right) \left(\widehat{q-r}^2\right)
      - 2\; \left(\widehat{p+k}^2\right) \left(\widehat{q-r}^2\right)
     \right] \; ,
\end{eqnarray}
from the black spot and 
\begin{eqnarray}
  - {1\over 8 g} \;  & &
{\sum}_p\;
{\sum}_q\;
{\sum}_k\;
{\sum}_r\; 
 L^2 \;\delta^2(p+q+k+r) \nonumber \\
 && \qquad \qquad \qquad \qquad \times \; \sum_{ab}  
     \pi^a(p) \, \pi^a(q) \, \pi^b(k) \, \pi^b(r) \, 
      \Pi_{p+q} \; ,
\end{eqnarray}
which is the vertex coming from the 0--loop Symanzik part of the
action. The contraction gives
\begin{equation}
 -g^3 \;{1\over 16} \; c_9 \left[ \left(n-1\right)^2 I_1 +
                    \left(n-1\right) \left( I_1 + I_2 \right) \right]
                    \; ,
\end{equation}
where 
\begin{eqnarray}
 I_1 &\equiv& 
{\sum}_p\;
{\sum}_q\;
{\sum}_k\;
{\sum}_r\;
 L^2 \;\delta^2(p+q+k+r) \nonumber \\
 && \qquad \qquad \qquad \qquad \times {{\Pi_{p+q} 
     \left[ \left(\widehat{p+k}^2\right)^2 +
            \left(\widehat{p-k}^2\right) \left(\widehat{q-r}^2\right)
      - 2\; \left(\widehat{p+k}^2\right) \left(\widehat{q-r}^2\right)
     \right]} \over \Pi_p \; \Pi_q \; \Pi_k \; \Pi_r } 
 \nonumber \\
 I_2 &\equiv& 
{\sum}_p\;
{\sum}_q\;
{\sum}_k\;
{\sum}_r\;
 L^2 \;\delta^2(p+q+k+r) \nonumber \\
 && \qquad \qquad \qquad \qquad \times {{\Pi_{p+q} 
     \left[ \left(\widehat{p+q}^2\right)^2 +
            \left(\widehat{p-q}^2\right) \left(\widehat{k-r}^2\right)
      - 2\; \left(\widehat{p+q}^2\right) \left(\widehat{k-r}^2\right)
     \right]} \over \Pi_p \; \Pi_q \; \Pi_k \; \Pi_r } \; .
\label{sunday31}
\end{eqnarray}
The numerators can be easily worked out. For instance, the square
brackets in the numerator of $I_1$ can be rewritten by using the Kronecker
delta in Eq.(\ref{sunday31}) ($\Delta_{p,q}$ is introduced in Appendix~A),
\begin{eqnarray}
 \left(\widehat{p+k}^2\right)^2 +
            \left(\widehat{p-k}^2\right) \left(\widehat{q-r}^2\right)
      &-& 2\; \left(\widehat{p+k}^2\right) \left(\widehat{q-r}^2\right)
   \nonumber \\
  &=&\; \left(\Delta_{p,k} + \widehat{p}^2 + \widehat{k}^2 \right) 
      \Big(\Delta_{q,r} + \widehat{q}^2 + \widehat{r}^2 \Big) 
     \nonumber \\
  && + \left(\Delta_{p,-k} + \widehat{p}^2 + \widehat{k}^2 \right)
      \Big(\Delta_{q,-r} + \widehat{q}^2 + \widehat{r}^2 \Big) 
     \nonumber \\
 &&  -2\; \left(\Delta_{p,k} + \widehat{p}^2 + \widehat{k}^2 \right)
      \Big(\Delta_{q,-r} + \widehat{q}^2 + \widehat{r}^2 \Big)
     \nonumber \\
  &=& \Delta_{p,k}\; \Delta_{q,r} + \Delta_{p,-k}\; \Delta_{q,-r}
      - 2\; \Delta_{p,k}\; \Delta_{q,-r} \; ,
\end{eqnarray}
which is true under the integration. After taking the limit
$L\rightarrow\infty$, this expression leads immediately
to the final result for  
$I_1=S_{11} + S_{14} - 2\; S_{15}$. Analogously 
$I_2=S_{10} + S_{16} - 2\; S_{17}$. The integrals $S_i$ are defined and
evaluated in Appendix~A.

\vskip 1cm

\begin{figure}[htbp]
\centerline{\epsfig{file=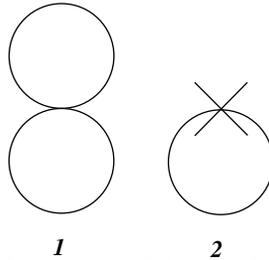,width=0.2\textwidth}}
\caption{Feynman diagrams contributing to the free energy of
the standard and 0--loop Symanzik actions at two loops. The lines
represent the propagation of the scalar field $\vec\pi$, the cross 
stands for a vertex from the measure action.}
\label{fig:1}
\end{figure}

\begin{figure}[htbp]
\centerline{\epsfig{file=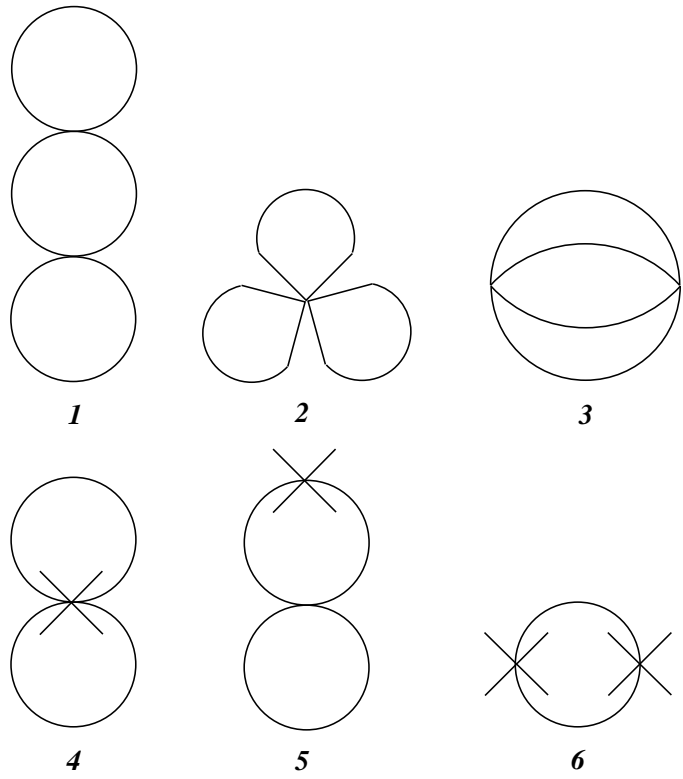,width=0.5\textwidth}}
\caption{The same as Fig.~\ref{fig:1} at three loops.}
\label{fig:2}
\end{figure}

\begin{figure}[htbp]
\centerline{\epsfig{file=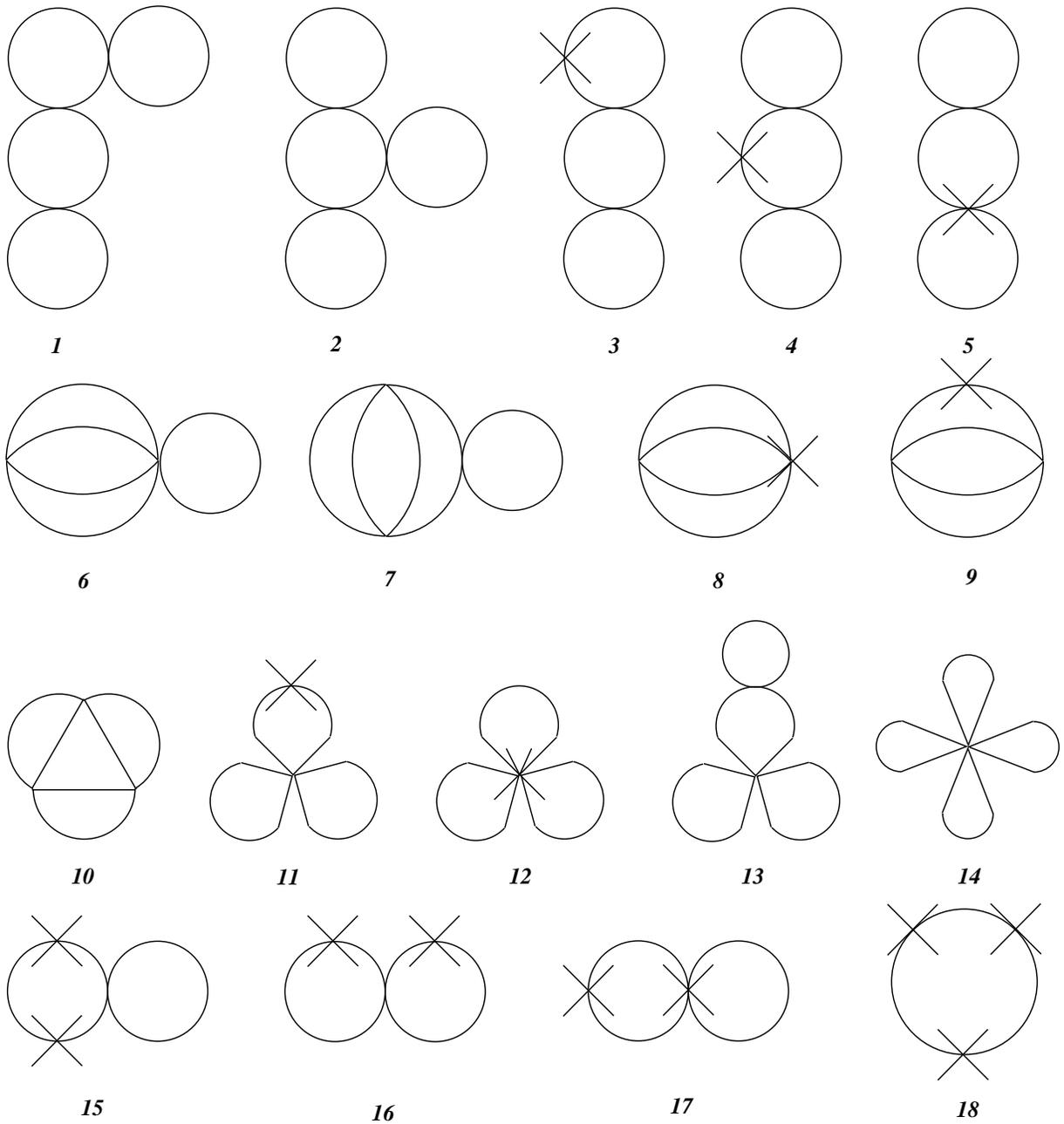,width=0.9\textwidth}}
\caption{The same as Fig.~\ref{fig:1} at four loops.}
\label{fig:3}
\end{figure}

\newpage

\begin{figure}[htbp]
\centerline{\epsfig{file=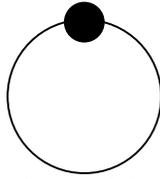,width=0.12\textwidth}}
\caption{The Feynman diagrams necessary for the calculation of the
free energy at two loops for the 1--loop Symanzik action are those 
shown in Fig.~\ref{fig:1} plus the diagram displayed in this figure.
The black spot indicates a vertex proportional to some coefficient
$c_i$.}
\label{fig:4}
\end{figure}

\begin{figure}[htbp]
\centerline{\epsfig{file=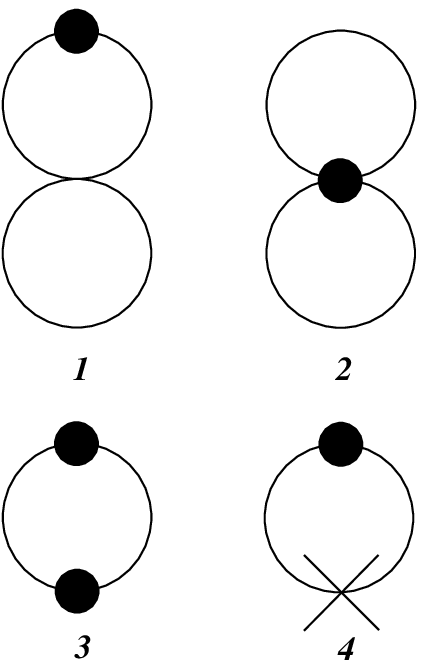,width=0.3\textwidth}}
\caption{The free energy at three loops for the 1--loop Symanzik action
is calculated by adding up the diagrams of Fig.~\ref{fig:2} to those displayed
in this figure. Same notation as in Fig.~\ref{fig:4}.}
\label{fig:5}
\end{figure}

\begin{figure}[htbp]
\centerline{\epsfig{file=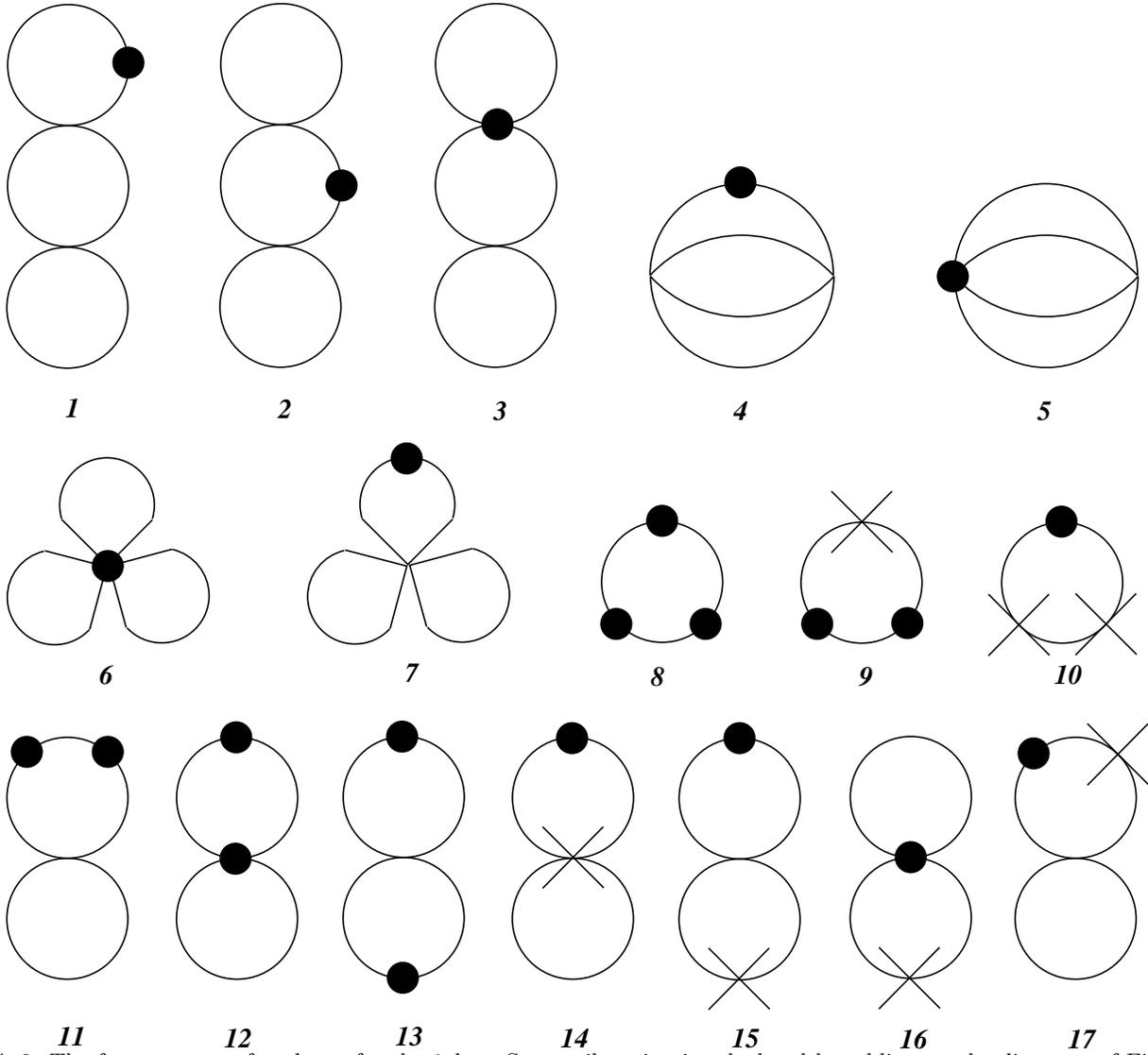,width=0.9\textwidth}}
\caption{The free energy at four loops for the 1--loop Symanzik action
is calculated by adding up the diagrams of Fig.~\ref{fig:3} to those displayed
in this figure. Same notation as in Fig.~\ref{fig:4}.}
\label{fig:6}
\end{figure}

\section{Results for $E$ and effective scheme}

In this section we will give the results for the internal energy $E$ and will
explain how to obtain the coefficients for $\beta^L$ in the corresponding
effective scheme. We will show the results for all three actions. Those
for the standard action are not new~\cite{bl,cp1,ml,abc} but we have 
checked all of them. We write the perturbative expansion of $E$ as
\begin{equation}
 E=w_0- w_1 \; g - w_2\; g^2 - w_3\; g^3 - w_4 \; g^4 - \cdots \;.
\end{equation}
Then,
\begin{eqnarray}
 w^{\rm standard}_0 &=& 1 \;, \nonumber \\
 w^{\rm standard}_1 &=& {\left(n-1\right)\over 4} \;,\nonumber \\
 w^{\rm standard}_2 &=& {\left(n-1\right) \over 32} \;,\nonumber \\
 w^{\rm standard}_3 &=& {\left(n-1\right)^2\over 16} K +
         {\left(n-1\right)\over 16} \left( {1\over 6} - K + 
         {1\over 3} J\right) \;,\nonumber \\
 w^{\rm standard}_4 &=& {3\over 8} \left(n-1\right)
   \left( {1\over 128} - {1\over 2} H_1 - {1\over 4} H_2 - {1\over 3} H_3 +
          {1\over 24} J - {1\over 8} K - {1\over 4} H_5 \right) \nonumber \\
  && +{3\over 8} \left(n-1\right)^2 
   \left( {1\over 256} + {1\over 2} H_1 + {1\over 4} H_2 + {1\over 3} H_3 +
          {1\over 12} H_4 + {1\over 8} K + {1\over 3} H_5 \right) \nonumber \\
   && -{\left(n-1\right)^3\over 32} H_5 \;.
\end{eqnarray}
The integrals $K$, $J$ and $H_i$ for $i=1,\dots\,,5$ are defined and
calculated in~\cite{cp1,abc}. They are shown in our Appendix~A. 

The results for the 0--loop Symanzik action are known up to 
three loops~\cite{dgfpv,abc} and
here we have added the fourth order. The first terms 
$w^{\rm 0-Symanzik}_0$, ..., $w^{\rm 0-Symanzik}_3$ have been 
checked obtaining full agreement. The complete set of coefficients is
\begin{eqnarray}
 w^{\rm 0-Symanzik}_0 &=& {15\over 12} \;, \nonumber \\
 w^{\rm 0-Symanzik}_1 &=& {\left(n-1\right)\over 4} \;, \nonumber \\
 w^{\rm 0-Symanzik}_2 &=& {\left(n-1\right)\over 48}Y_1 
                          \left(1 - {5\over 24} Y_1\right) 
       \;, \nonumber \\
 w^{\rm 0-Symanzik}_3 &=& {\left(n-1\right)^2\over 16} K^S +
         {\left(n-1\right)\over 16} \Bigg( {1\over 3} - K^S + 
         {1\over 3} J^S +{1\over 36} Y_2 \nonumber \\
 && \qquad -Y_1\left({5\over 12} + {5\over 216} Y_2\right) +
    {Y_1}^2 \left( {11\over 48} + {25\over 5184} Y_2\right) -
    {205\over 5184}{Y_1}^3\Bigg) \;,\nonumber \\
 w^{\rm 0-Symanzik}_4 &=&  
   { \left(n-1\right)}\,\Bigg( -{5\over {256}} - {1\over {96}}S_9 - 
      {{3}\over {16}}H_1^S - {{3}\over {32}}H_2^S - 
    {1\over 8}H_3^S + {{1}\over {16}}J^S - {1\over 8} \overline{J^S}+ 
      {1\over {16}} \widetilde{J^S} \nonumber \\
 && \qquad\qquad -{{3}\over {16}} K^S+ 
      {{3}\over 8}\overline{K^S} - {{3}\over {16}}\widetilde{K^S} - 
      {{3}\over {32}}H_5^S + {{35}\over {768}}Y_1 + 
      {{5}\over {1152}}S_9\,Y_1 - 
      {{1}\over {48}}J^S Y_1 \nonumber \\
 && \qquad\qquad+ {{5}\over {96}}\overline{J^S}\,Y_1 - 
      {{5}\over {192}}\widetilde{J^S}\,{ Y_1} + 
      {{1}\over {16}}{ K^S}\,{ Y_1} - 
      {{5}\over {32}} { \overline{K^S}}\,{ Y_1}+ 
      {{5}\over {64}}{\widetilde{K^S}}\,{ Y_1} - 
      {{75}\over {2048}}{Y_1}^2 \nonumber \\
 && \qquad\qquad    + {{53}\over {4096}}{{ Y_1}}^3 - 
      {{9005}\over {5308416}}{{ Y_1}}^4 - {{5}\over {1536}} Y_2 + 
      {{91}\over {18432}}{ Y_1}\,{ Y_2} - 
      {{535}\over {221184}}{{ Y_1}}^2\,{ Y_2} \nonumber \\
&&\qquad\qquad+   {{1025}\over {2654208}}{{{ Y_1}}^3}\,{ Y_2} - 
      {{5}\over {55296}}{{ Y_2}}^2 
   +  {{25}\over {331776}}{ Y_1}\,{{{ Y_2}}^2} 
   -  {{125}\over {7962624}} {{ Y_1}}^2\,{{{ Y_2}}^2} \nonumber \\
&& \qquad\qquad +  {1\over {6912}} Y_3- {{5}\over {27648}}{ Y_1}\,{ Y_3} + 
      {{25}\over {331776}}{{ Y_1}}^2\,{ Y_3} - 
      {{125}\over {11943936}}{{{ Y_1}}^3}\,{ Y_3} \Bigg) \nonumber \\
 && +{{{ \left(n-1\right)}}^2}\,\Bigg( {5\over {512}} + {1\over {96}}S_9 + 
      {{3}\over {16}}H_1^S + {{3}\over {32}}H_2^S + 
      {1\over 8}H_3^S + {{3}\over {16}}K^S - 
      {{3}\over 8}\overline{K^S} + {{3}\over {16}} \widetilde{K^S}
 \nonumber \\
&&\qquad\qquad  +{1\over {32}}H_4^S + {1\over 8}H_5^S - {{25}\over {1536}}Y_1 - 
      {{5}\over {1152}} S_9 \,Y_1- 
      {{1}\over {16}}K^S \,Y_1 + 
      {{5}\over {32}}{ \overline{K^S}}\,{ Y_1} \nonumber \\
&& \qquad\qquad -   {{5}\over {64}} { \widetilde{K^S}}\,{ Y_1}+ 
      {{125}\over {12288}}{{ Y_1}}^2 - 
      {{593}\over {221184}}{{ Y_1}}^3 + 
      {{2725}\over {10616832}} {{ Y_1}}^4\Bigg)\nonumber \\
&& -  { \left(n-1\right)^3\over 32} H_5^S \;.
\end{eqnarray}
The integrals $K^S$, $J^S$, $\overline{K^S}$, $\overline{J^S}$,
$\widetilde{K^S}$, $\widetilde{J^S}$, $Y_1$, $Y_2$  and $Y_3$ were
introduced in~\cite{abc}. They are shown for completeness in Appendix~A.
The new integrals are $S_9$, $H^S_1$, $H^S_2$, $H^S_3$, $H^S_4$ and
$H^S_5$. All of them are listed and evaluated in the Appendix.

The expression of the internal energy for the 1--loop Symanzik improved 
action is a new result of the present paper for all loops 
and is written in terms of the following coefficients
\begin{eqnarray}
 w^{\rm 1-Symanzik}_0 &=& w^{\rm 0-Symanzik}_0 \;,\nonumber\\
 w^{\rm 1-Symanzik}_1 &=& w^{\rm 0-Symanzik}_1 \;,\nonumber\\
 w^{\rm 1-Symanzik}_2 &=& w^{\rm 0-Symanzik}_2 + 
    {\left(n-1\right) \over 2} \left(c_5 \left(3 + {1\over 144} Y_{2,1}\right)
    + c_6 Y_1\right) \;, \nonumber \\
 w^{\rm 1-Symanzik}_3 &=& w^{\rm 0-Symanzik}_3 + 
    \left(n-1\right)^2 \left(\left(c_7+ {1\over 2}c_8\right)
    \left(1 - {1\over 6}Y_1 + {1\over 144} {Y_1}^2\right) +
    c_9 \left({1\over 2} -{1\over 3} Y_1 + {1\over 18} {Y_1}^2\right)\right) 
   \nonumber \\
   && \qquad +\left(n-1\right) \Bigg(
       c_9 \left({3\over 2} - Y_1 +{1\over 6} {Y_1}^2\right) +
       c_8 \left(1 - {2\over 3} Y_1 + {25\over 144} {Y_1}^2\right) 
   \nonumber \\
   && \qquad \qquad \qquad + c_7 \left( {25\over 16} - {25\over 24} Y_1 +
       {17\over 72} {Y_1}^2 + {1\over 384} Y_{2,1} -
    {1\over 1152} Y_1 \,Y_{2,1} + {1\over 331776} \left(Y_{2,1}\right)^2\right) 
   \nonumber \\
   && \qquad \qquad \qquad +c_6^2 \,Y_2 + c_6\left(
     {3\over 2} - {5\over 4} Y_1 +{29\over 96} {Y_1}^2 +
     {1\over 12} Y_2 - {5\over 144}Y_1\,Y_2\right)
   \nonumber \\
   && \qquad \qquad \qquad +c_5\,\bigg({33\over 16} -{5\over 4}Y_1 +
       {35\over 144} {Y_1}^2 - {17\over 1152}Y_{2,1} + 
      {5\over 864} Y_1\,Y_{2,1} +{1\over 331776}\left(Y_{2,1}\right)^2 
   \nonumber \\
   && \qquad\qquad\qquad +
      {1\over 1728} Y_{3,2} - {5\over 20736} Y_1\,Y_{3,2}\bigg) +
      c_5\,c_6\,\left(24 - {1\over 3} Y_{2,1} + {1\over 72} Y_{3,2}\right)
   \nonumber \\
   && \qquad  \qquad  \qquad 
      + c_5^2\left({55\over 4} - {1\over 432}Y_{3,1} + {1\over 20736}
       Y_{4,2}\right)\;\;\Bigg) \;.
\end{eqnarray}
The analytical form of $w^{\rm 1-Symanzik}_4$ is very lengthy and 
its explicit expression is reported in Appendix~D.

Numerically these expansions read (we only report five significant
digits, which are usually enough for simulation purposes, 
although the results shown in Appendix~A for the finite integrals 
can be readily calculated with greater precision)
\begin{eqnarray}
 E^{\rm standard} &=& 1 - {n-1\over 4}\; g - {n-1\over 32 }\;g^2 -
   \left[0.0072699 \left(n-1\right) + 0.0059930 \left(n-1\right)^2\right]
   \;g^3 \nonumber \\ 
  && - \left[0.0028167 \left(n-1\right) + 0.0034299 \left(n-1\right)^2 +
         0.0015673 \left(n-1\right)^3\right] \;g^4 \;\;,\nonumber \\
  && \nonumber \\ 
 E^{\rm 0-Symanzik} &=& {15\over 12} - {n-1\over 4}\; g - 0.024449\;
      \left(n-1\right)\;g^2 -
   \left[0.0044905 \left(n-1\right) + 0.0042082 \left(n-1\right)^2\right]
   \;g^3 \nonumber \\ 
  && - \left[0.0014508 \left(n-1\right) +0.0017541 \left(n-1\right)^2 +
         0.0010241 \left(n-1\right)^3\right]\;g^4 \;\;,\nonumber \\
  && \nonumber \\ 
 E^{\rm 1-Symanzik} &=& {15\over 12} - {n-1\over 4}\; g - 
  \Big[\left(0.024449 + 1.60443 \; c_5 + 1.02179 \; c_6\right)
   \left(n-1\right) \Big] \; g^2 
   \nonumber \\
 && - \Big[\big( 0.0044905 + 12.5286 \;c_5^2  + 0.26629 \;c_6 + 4.7831 \;c_6^2 + 
      0.44431\; c_5 + 15.0482 \;c_5 \;c_6 \nonumber \\
 && \qquad\qquad + 0.44752 \;c_7 + 0.36265 \;c_8 + 0.15246 \;c_9 
     \big) \left(n-1\right) \nonumber \\ 
 &&  \qquad\qquad + \left( 0.0042082 +
    0.68841 \;c_7 + 0.34420 \;c_8 + 0.050819 \;c_9\right)
   \left(n-1\right)^2\Big]
   \;g^3 \nonumber \\ 
  && - \Big[\big( 0.0014508 +
  0.11759 \;c_5 + 4.7668 \;c_5^2  + 108.382 \;c_5^3  
   + 0.068884 \;c_6 + 5.8854 \;c_5 \;c_6 
      \nonumber \\
 && \qquad\qquad + 186.142 \;c_5^2  \;c_6 + 
    1.7585 \;c_6^2  + 111.566 \;c_5 \;c_6^2  + 
  23.6332 \;c_6^3  + 0.11683 \;c_7 
      \nonumber \\ 
 && \qquad\qquad + 9.2935 \;c_5 \;c_7 + 5.7226 \;c_6 \;c_7 +
  0.078923 \;c_8 + 7.1066 \;c_5 \;c_8 + 4.5248 \;c_6 \;c_8 
      \nonumber \\
 && \qquad\qquad - 0.031694 \;c_9 + 
  2.0109 \;c_5 \;c_9 + 1.2889 \;c_6 \;c_9 \big) \left(n-1\right) 
      \nonumber \\
 && \qquad\qquad+ \big( 0.0017541 + 0.068786 \;c_5 + 0.047639 \;c_6 + 
      0.15515 \;c_7 + 12.853 \;c_5 \;c_7 
      \nonumber \\
 && \qquad\qquad +
  8.18909 \;c_6 \;c_7 + 0.069749 \;c_8 + 6.4265 \;c_5 \;c_8 + 4.0946 \;c_6 \;c_8
      \nonumber \\
 && \qquad\qquad -  0.011464 \;c_9 + 0.67029 \;c_5 \;c_9 + 0.42964 \;c_6 \;c_9
                  \big)\left(n-1\right)^2 
      \nonumber \\
 && \qquad\qquad +  0.0010241 \left(n-1\right)^3\Big]\;g^4 \;\;.
\label{numericale}
\end{eqnarray}
Notice that at order $O(g)$ the result 
is the same in all three cases. This is a 
consequence of the equipartition of the energy. The numerical value of
the quartic coefficient in $E^{\rm standard}$ differs roughly by 3\%
from the result in Ref.~\cite{abc}. This is due to the more accurate
determination of the integral $H_3$ obtained in the present paper 
(see Appendix~A). This difference is too small to change at all any
conclusion of Ref.~\cite{abc}.

Now the energy--based effective coupling constant $g_E$ is defined 
in a non--perturbative way,
\begin{equation}
 g_E \equiv {w_0 - E^{\rm MC} \over w_1}\;,
\end{equation}
where $E^{\rm MC}$ is the Monte Carlo 
measured value of the internal energy at some
value of the bare coupling $g$. The perturbative expansion 
of $g_E$ in terms of this 
coupling $g$ is obtained from the previously calculated expansion for $E$
\begin{equation}
 g_E = g + {w_2 \over w_1} g^2 + {w_3 \over w_1} g^3 + {w_4 \over w_1} g^4
 + \cdots\;.
\label{expansione}
\end{equation}
The lattice beta function $\beta^L$ in terms of the bare coupling $g$ 
is~\cite{ap}
\begin{equation}
 \beta^L(g)\equiv -a {\hbox{d} g\over \hbox{d}a} = 
   -\beta_0 g^2 -\beta_1 g^3 -\beta_2 g^4 -\beta_3 g^5 - \cdots\;,
\label{beta}
\end{equation}
and in terms of $g_E$ it becomes
\begin{eqnarray}
 \beta^L(g_E) &=& \beta^L\left(g\left(g_E\right)\right) \;\;
                {\hbox{d} g_E\over \hbox{d}g}\left(g_E\right) \nonumber \\
  &=& -\beta_0\; g_E^2 -\beta_1\; g_E^3 -
     {{\beta_2 \; w_1^2 - \beta_1\; w_1 w_2 - \beta_0 \left( w_2^2 -
      w_1\, w_3 \right)} \over w_1^2}\; g_E^4 \nonumber \\
  && - {{\beta_3\; w_1^3- 2\, \beta_2\; w_1^2\; w_2+ \beta_1\; w_1\; w_2^2 +
     2\, \beta_0\; \left( 2 \,w_2^3 - 3\, w_1\; w_2\; w_3 + 
      w_1^2\; w_4\right)} \over
         w_1^3}\; g_E^5 - \cdots \;,
\end{eqnarray}
where the function $g(g_E)$ is obtained by inverting Eq.(\ref{expansione}).

 The integration of the beta function yields the dependence of the 
lattice spacing $a$ on the coupling constant 
\begin{equation} 
 a\Lambda = \left(\beta_0 g\right)^{-\beta_1/\beta_0^2}
            \exp\left(-{1\over \beta_0 g}\right) \left(1 + O(g)\right) \;,
\label{alambda}
\end{equation}
where $\Lambda$ is the integration constant, the lattice Lambda
parameter. An analogous equation can be derived
in the effective scheme with an integration constant $\Lambda_E$. From 
Eq.(\ref{expansione}) and Eq.(\ref{alambda}) the ratio of these two constants
can be exactly determined,
\begin{equation}
 \Lambda_E = \Lambda \exp \left\{ {w_2\over w_1\;\beta_0}\right\} \;.
\end{equation}

\section{Conclusions}

 We have calculated the perturbative expansion of the internal energy 
for the tree--level and 1--loop improved Symanzik actions on the
lattice for the 2D nonlinear $\sigma$--model with symmetry
$O(n)$ up to fourth order in the coupling constant. 
These results are shown in Eq.(\ref{numericale}).
The definitions that we have adopted for the internal energy $E$ are shown
in Eq.(\ref{representing}). These expansions allow the definition 
of an effective coupling $g_E$. We expect that the series which determine
the lattice spacing in terms of the coupling constant are better behaved
if expressed in powers of $g_E$. This hope will be checked in a future
publication where we plan to calculate the mass gap of the model
through a Monte Carlo simulation. To this end, we have also given the
analytic expression of the lattice beta function in terms of $g_E$. 

 The calculation is rather involved and it has been done separately by 
the two authors. Only the final results were compared. Also the numerical
value of the integrals in Appendix~A has been obtained independently
and checked afterwards. Further checks were done: for example some
subsets of diagrams must altogether yield an
IR--finite result. These tests
come out when we consider other definitions for the
energy operator different from the one shown in
Eq.(\ref{representing}). For instance if we use the whole expression
of the 1--loop Symanzik action as an energy operator,
Eq.(\ref{s1loop}), then diagrams 9, 11, 12 and 13 of Figure~6 are
multiplied by a different factor from the rest of diagrams. Then this
subset, taken separately, must produce an IR--finite result.

\vskip 1cm

\newpage

{\centerline{\bf Acknowledgements}}

\vskip 2mm

It is a pleasure to thank Paolo Butera, Marco Comi and Giuseppe
Marchesini for useful discussions, Andrea Pelissetto for 
clarifying comments on the IR regularization
and Paolo Butera for a critical reading of the manuscript. 

\vskip 1cm

\section{Appendix~A}
{\centerline {\bf List of finite integrals}}
\vskip 5mm

We will give the list of finite integrals that we have used to express
the results for $E$ as a manifestly finite quantity.
The basic notation is introduced in section~3. Besides, the following
definitions will be needed:
\begin{eqnarray}
\Delta_{p,q} &\equiv& \widehat{(p+q)}^2 - 
   \widehat{p}^2 - \widehat{q}^2 \;,\nonumber \\
\Delta_{p,-q} &\equiv& \widehat{(p-q)}^2 -
   \widehat{p}^2 - \widehat{q}^2 \;,\nonumber \\
\Delta^S_{p,q} &\equiv& \Pi_{p+q} - \Pi_p -\Pi_q \;,\nonumber \\
\Delta^S_{p,-q} &\equiv& \Pi_{p-q} - \Pi_p -\Pi_q \;,\nonumber \\
\Delta^\Box_{p,q} &\equiv& \Box_{p+q} - \Box_p -\Box_q \;,\nonumber \\
\Delta^\mu_{p,q} &\equiv& \widehat{(p+q)}^2_\mu - \widehat{p}^2_\mu - 
  \widehat{q}^2_\mu \;.
\end{eqnarray}
In the three--loop integrals we use the notation
\begin{equation}
\int D_3 \equiv 
\int_{-\pi}^{+\pi}  {{{\rm d^2} p} \over  \left(2 \pi\right)^2}
\int_{-\pi}^{+\pi}  {{{\rm d^2} q} \over  \left(2 \pi\right)^2}
\int_{-\pi}^{+\pi}  {{{\rm d^2} k} \over  \left(2 \pi\right)^2}
\int_{-\pi}^{+\pi}  {{{\rm d^2} r} \over  \left(2 \pi\right)^2}
\left(2 \pi\right)^2 \delta^2(p+q+k+r) \;,
\label{defD3}
\end{equation}
and the definitions and numerical values of the integrals are
\begin{equation}
 K \equiv \int D_3 \;\; {\Delta_{p,q}\;\Delta_{k,r}\over {
       \widehat{p}^2\;\widehat{q}^2\;\widehat{k}^2\;\widehat{r}^2}} =
       0.0958876\;,
\end{equation}
\begin{equation}
 J \equiv \int D_3 \;\;  {\left(\Sigma_{pqkr}\right)^2 \over {
       \widehat{p}^2\;\widehat{q}^2\;\widehat{k}^2\;\widehat{r}^2}} =
       0.136620\;,
\end{equation}
\begin{equation}
 K^S \equiv \int D_3 \;\;  {\Delta^S_{p,q}\;\Delta^S_{k,r}\over {
       \Pi_p\;\Pi_q\;\Pi_k\;\Pi_r}} = 0.0673313\;,
\end{equation}
\begin{equation}
 J^S \equiv \int D_3  \;\; {\left(\Sigma^S_{pqkr}\right)^2 \over {
       \Pi_p\;\Pi_q\;\Pi_k\;\Pi_r}} =
       0.104551\;,
\end{equation}
\begin{equation}
 \overline{K^S} \equiv \int D_3  \;\; {\Delta^S_{p,q}\;
                 \Delta^S_{k,r}\;\widehat{p}^2\over {
       \left(\Pi_p\right)^2\;\Pi_q\;\Pi_k\;\Pi_r}} =
       0.0572726\;,
\end{equation}
\begin{equation}
 \overline{J^S}\equiv\int D_3 \;\; 
       {\left(\Sigma^S_{pqkr}\right)^2\;\widehat{p}^2\over{
       \left(\Pi_p\right)^2\;\Pi_q\;\Pi_k\;\Pi_r}} =
       0.0867807\;,
\end{equation}
\begin{equation}
 \widetilde{K^S} \equiv \int D_3  \;\; {\Delta^S_{p,q}\;\Delta_{k,r}\over {
       \Pi_p\;\Pi_q\;\Pi_k\;\Pi_r}} =
       0.0578002\;,
\end{equation}
\begin{equation}
 \widetilde{J^S} \equiv \int D_3 \;\;  {\Sigma^S_{pqkr}\;\Sigma_{pqkr} \over {
       \Pi_p\;\Pi_q\;\Pi_k\;\Pi_r}} =
       0.0809553\;,
\end{equation}
\begin{equation}
S_1 \equiv \int D_3  \;\;
 {{\Delta^S_{p,q} \left(\Delta_{k,r}\right)^2} \over
  \Pi_p\;\Pi_q\; \Pi_k\; \Pi_r } = -0.283407\;,
\end{equation}
\begin{equation}
S_2 \equiv \int D_3  \;\;
 {{\Delta^S_{p,q} \;\Delta_{k,r}\; \widehat{k}^2} \over
  \Pi_p \;\Pi_q \;\Pi_k \;\Pi_r }=0.184636\;,
\end{equation}
\begin{equation}
S_3 \equiv \int D_3  \;\;
 {{\Delta^S_{p,q} \;\widehat{k}^2\; \widehat{r}^2} \over
  \Pi_p \;\Pi_q \;\Pi_k \;\Pi_r }=-0.134904\;,
\end{equation}
\begin{equation}
S_4 \equiv \int D_3  \;\;
 {{\Sigma_{pqkr}^S \;\left(\widehat{p}^2\right)^2 } \over
  \Pi_p \;\Pi_q \;\Pi_k \;\Pi_r }=0.148440\;,
\end{equation}
\begin{equation}
S_5 \equiv \int D_3  \;\;
 {{\Sigma_{pqkr}^S \;\Delta_{p,q} \;\Delta_{k,r}} \over
  \Pi_p \;\Pi_q \;\Pi_k \;\Pi_r }=0.261935\;,
\end{equation}
\begin{equation}
S_6 \equiv \int D_3  \;\;
 {{ \Sigma_{pqkr}^S \;\Delta_{p,q}\; \widehat{k}^2} \over
  \Pi_p \;\Pi_q \;\Pi_k \;\Pi_r }=-0.181963\;,
\end{equation}
\begin{equation}
S_7 \equiv \int D_3  \;\;
 {{ \Sigma_{pqkr}^S \;\widehat{k}^2\; \widehat{r}^2} \over 
  \Pi_p \;\Pi_q \;\Pi_k\; \Pi_r }=0.124096\;,
\end{equation}
\begin{equation}
S_8 \equiv \int D_3  \;\;
 {{\Delta^S_{p,q}\; \Delta^\Box_{k,r}} \over
  \Pi_p \;\Pi_q \;\Pi_k\; \Pi_r }=0.114374\;,
\end{equation}
\begin{equation}
S_9 \equiv \int D_3  \;\;
 {{\Sigma_{pqkr}^S \;\Box_p} \over
  \Pi_p \;\Pi_q \;\Pi_k \;\Pi_r }=0.0891356\;,
\end{equation}
\begin{equation}
S_{10} \equiv \int D_3  \;\;
 {{\Pi_{p+q} \;\Delta_{p,q}\; \Delta_{k,r}} \over
  \Pi_p \;\Pi_q \;\Pi_k \;\Pi_r }=0.0881094\;,
\end{equation}
\begin{equation}
S_{11} \equiv \int D_3  \;\;
 {{\Pi_{p+q}\; \Delta_{p,k}\; \Delta_{q,r}} \over
  \Pi_p \;\Pi_q \;\Pi_k \;\Pi_r }=0.215207\;,
\end{equation}
\begin{equation}
S_{12} \equiv \int D_3  \;\;
 {{\Pi_{p+q}\; \sum_\mu \Delta_{p,q}^\mu\, \Delta_{k,r}^\mu} \over
  \Pi_p \;\Pi_q \;\Pi_k\; \Pi_r }=0.0649188\;,
\end{equation}
\begin{equation}
S_{13} \equiv \int D_3  \;\;
 {{\Pi_{p+q} \;\sum_\mu \Delta_{p,k}^\mu\, \Delta_{q,r}^\mu} \over
  \Pi_p \;\Pi_q \;\Pi_k \;\Pi_r }=0.113020\;,
\end{equation}
\begin{equation}
S_{14} \equiv \int D_3  \;\;
 {{\Pi_{p+q}\; \Delta_{p,-k}\; \Delta_{q,-r}} \over
  \Pi_p \;\Pi_q \;\Pi_k\; \Pi_r }=0.170225\;,
\end{equation}
\begin{equation}
S_{15} \equiv \int D_3  \;\;
 {{\Pi_{p+q}\; \Delta_{p,k}\; \Delta_{q,-r}} \over
  \Pi_p \;\Pi_q \;\Pi_k \;\Pi_r }=0.108536\;,
\end{equation}
\begin{equation}
S_{16} \equiv \int D_3  \;\;
 {{\Pi_{p+q}\; \Delta_{p,-q}\; \Delta_{k,-r}} \over
  \Pi_p \;\Pi_q \;\Pi_k\; \Pi_r }=0.233959\;,
\end{equation}
\begin{equation}
S_{17} \equiv \int D_3  \;\;
 {{\Pi_{p+q}\; \Delta_{p,q}\; \Delta_{k,-r}} \over
  \Pi_p \;\Pi_q\; \Pi_k \;\Pi_r }=0.0963352\;,
\end{equation}
\begin{equation}
S_{18} \equiv \int D_3  \;\; {
 \Delta^S_{p,q} \;\Delta^S_{k,r}\; \left( \widehat{p}^2 \right)^2
 \over  \left(\Pi_p\right)^2 \;\Pi_q \;\Pi_k \;\Pi_r }=0.184379\;,
\end{equation}
\begin{equation}
S_{19} \equiv \int D_3  \;\; {
  \Delta^S_{p,q} \;\Delta^S_{k,r} \;\Box_p
 \over  \left(\Pi_p\right)^2\; \Pi_q\; \Pi_k\; \Pi_r }=0.120705\;,
\end{equation}
\begin{equation}
S_{20} \equiv \int D_3  \;\; {
 \left(\Sigma_{pqkr}^S\right)^2 \;\left( \widehat{p}^2 \right)^2
 \over  \left(\Pi_p\right)^2\; \Pi_q\; \Pi_k\; \Pi_r }=0.335775\;,
\end{equation}
\begin{equation}
S_{21} \equiv \int D_3  \;\; {
  \left(\Sigma_{pqkr}^S\right)^2 \;\Box_p
 \over  \left(\Pi_p\right)^2 \;\Pi_q \;\Pi_k\; \Pi_r }=0.213243\;.
\end{equation}

The above listed integrals are not all independent. 
In Appendix~B we give a few identities that these integrals satisfy.

The measure for the four--loop integrals is
\begin{eqnarray}
\int D_4 &\equiv& 
\int_{-\pi}^{+\pi}  {{{\rm d^2} p} \over  \left(2 \pi\right)^2}
\int_{-\pi}^{+\pi}  {{{\rm d^2} q} \over  \left(2 \pi\right)^2}
\int_{-\pi}^{+\pi}  {{{\rm d^2} k} \over  \left(2 \pi\right)^2}
\int_{-\pi}^{+\pi}  {{{\rm d^2} r} \over  \left(2 \pi\right)^2}
\int_{-\pi}^{+\pi}  {{{\rm d^2} s} \over  \left(2 \pi\right)^2}
\int_{-\pi}^{+\pi}  {{{\rm d^2} t} \over  \left(2 \pi\right)^2}\nonumber \\
&& \qquad \qquad \qquad \qquad \qquad \qquad \qquad \qquad \qquad \times
\left(2 \pi\right)^2 \delta^2(p+q+k+r) 
\left(2 \pi\right)^2 \delta^2(k+r+s+t) \;,
\end{eqnarray}
and the numerical values of the four--loop integrals are 
(the result of $H_3$ differs in the third significant digit from 
the less accurate result given in~\cite{abc})
\begin{equation}
 H_1 \equiv \int D_4 \;\;
 {\Delta_{p,q} \; \Delta_{k,r} \; \Sigma_{pqst} \over 
  \widehat{p}^2 \; \widehat{q}^2 \; \widehat{k}^2 \; 
  \widehat{r}^2 \; \widehat{s}^2 \;\widehat{t}^2 } = 0.0378134\;,
\label{h1}
\end{equation}
\begin{equation}
 H_2 \equiv \int D_4 \;\;
 {\Delta_{k,r}  \;\Sigma_{pqkr} \; \Sigma_{pqst} \over 
  \widehat{p}^2 \; \widehat{q}^2 \; \widehat{k}^2 \; 
  \widehat{r}^2 \; \widehat{s}^2 \; \widehat{t}^2 } = -0.0322778\;,
\end{equation}
\begin{equation}
 H_3 \equiv \int D_4 \;\;
 {\Delta_{p,k} \; \Delta_{r,s} \; \Delta_{q,-t} \over 
  \widehat{p}^2 \; \widehat{q}^2 \; \widehat{k}^2 \; 
  \widehat{r}^2 \; \widehat{s}^2 \; \widehat{t}^2 } = -0.0128736\;,
\label{h3}
\end{equation}
\begin{equation}
 H_4 \equiv \int D_4 \;\;
 {\Sigma_{pqkr} \; \Sigma_{krst} \; \Sigma_{pqst} \over 
  \widehat{p}^2 \; \widehat{q}^2 \; \widehat{k}^2 \; 
  \widehat{r}^2 \; \widehat{s}^2 \; \widehat{t}^2 } = 0.0411085\;,
\end{equation}
\begin{equation}
 H_5 \equiv \int D_4 \;\;
 {\Delta_{p,q} \; \Delta_{k,r} \; \Delta_{s,t} \over 
  \widehat{p}^2 \; \widehat{q}^2 \; \widehat{k}^2 \; 
  \widehat{r}^2 \; \widehat{s}^2 \; \widehat{t}^2 } = -0.0501531\;,
\end{equation}
\begin{equation}
 H^S_1 \equiv \int D_4 \;\;
 {\Delta^S_{p,q} \; \Delta^S_{k,r} \; \Sigma^S_{pqst} \over 
  \Pi_p \; \Pi_q \; \Pi_k \; 
  \Pi_r \; \Pi_s \;\Pi_t } = 0.0218345\;,
\end{equation}
\begin{equation}
 H^S_2 \equiv \int D_4 \;\;
 {\Delta^S_{k,r}  \;\Sigma^S_{pqkr} \; \Sigma^S_{pqst} \over 
  \Pi_p \; \Pi_q \; \Pi_k \; 
  \Pi_r \; \Pi_s \; \Pi_t } = -0.0181139\;,
\end{equation}
\begin{equation}
 H^S_3 \equiv \int D_4 \;\;
 {\Delta^S_{p,k} \; \Delta^S_{r,s} \; \Delta^S_{q,-t} \over 
  \Pi_p \; \Pi_q \; \Pi_k \; 
  \Pi_r \; \Pi_s \; \Pi_t } = -0.0042338\;,
\end{equation}
\begin{equation}
 H^S_4 \equiv \int D_4 \;\;
 {\Sigma^S_{pqkr} \; \Sigma^S_{krst} \; \Sigma^S_{pqst} \over 
  \Pi_p \; \Pi_q \; \Pi_k \; 
  \Pi_r \; \Pi_s \; \Pi_t } = 0.0262036\;,
\end{equation}
\begin{equation}
 H^S_5 \equiv \int D_4 \;\;
 {\Delta^S_{p,q} \; \Delta^S_{k,r} \; \Delta^S_{s,t} \over 
  \Pi_p \; \Pi_q \; \Pi_k \; 
  \Pi_r \; \Pi_s \; \Pi_t } = -0.0327709\;.
\label{hs5}
\end{equation}

On the other hand the one--loop integrals are defined as
\begin{equation}
 Y_i \equiv \int_{-\pi}^{+\pi} {\hbox{d}^2 q \over \left(2\pi\right)^2}
 \left( {\Box_q \over \Pi_q} \right)^i \;,
 \qquad\qquad\qquad
 Y_{i,j} \equiv \int_{-\pi}^{+\pi} {\hbox{d}^2 q \over \left(2\pi\right)^2}
  {\left(\Box_q\right)^i \over \left(\Pi_q\right)^j} \;,\qquad
 (2i \geq j) \;,
\end{equation}
and their results are listed in Table~1. Notice that $Y_{i,i}\equiv Y_i$.

\vskip 1cm

{\centerline {Table~1: One--loop integrals. }}
\vskip 2mm
{\centerline{
\begin{tabular}{|p{1cm}|p{4cm}|}
\hline
$Y_1$ & $\;$     2.0435764382979844236     \\ \hline
$Y_2$ & $\;$     4.7830710733439886212     \\ \hline
$Y_3$ & $\;$     11.816615246907788250     \\ \hline
$Y_{1,2}$ & $\;$ 0.4729502261432961899     \\ \hline
$Y_{2,1}$ & $\;$ 30.077096804291341057     \\ \hline
$Y_{3,1}$ & $\;$ 558.65986413777280387     \\ \hline
$Y_{3,2}$ & $\;$ 77.324121011413132160     \\ \hline
$Y_{4,1}$ & $\;$ 11817.841483609309517     \\ \hline
$Y_{4,2}$ & $\;$ 1489.1480965521674895     \\ \hline
$Y_{4,3}$ & $\;$ 202.26364872706189510     \\ \hline
$Y_{5,2}$ & $\;$ 32200.496224041766111     \\ \hline
$Y_{5,3}$ & $\;$ 4006.2729031961906982     \\ \hline
$Y_{6,3}$ & $\;$ 88276.902118545681915     \\ \hline
\end{tabular}} 
}
\vskip 5mm
Some of the one--loop integrals were introduced in Ref.~\cite{abc}.
All $Y_{i,j}$ with $i\not= j$ show up only in the results
for the 1--loop Symanzik action.

In the perturbative expansion of the 1--loop Symanzik action there appear
some vertices with a very high mass dimension. Once these vertices
are inserted in the corresponding Feynman diagrams, all propagators may
cancel leading to non--fractional integrands. In this case the following
expression can be useful
\begin{equation}
 \int_{-\pi}^{+\pi} {\hbox{d} q \over 2\pi} \left(\widehat{q}_\mu\right)^{2m} =
 2\; {2 m -1 \choose m }\;, \qquad\qquad m\geq 1\;.
\end{equation}
For example,
\begin{equation}
 \int_{-\pi}^{+\pi} {\hbox{d}^2 q \over \left(2\pi\right)^2}
  \;\widehat{q}^2 = 4  \;,
 \qquad\qquad
 \int_{-\pi}^{+\pi} {\hbox{d}^2 q \over \left(2\pi\right)^2}
  \;\Box_q = 12  \;,
 \qquad \qquad
  \int_{-\pi}^{+\pi} {\hbox{d}^2 q \over \left(2\pi\right)^2}
  \;\Pi_q = 5  \;.
\end{equation}

\section{Appendix~B}
{\centerline {\bf Identities among the integrals}}
\vskip 5mm

Some of the integrals $S_i$ in the above Appendix are actually related
among themselves. We have not taken advantage of these relationships in the
final expressions of section~4 because we are not sure to have discovered
all of them. In this Appendix we show the identities 
that we have found out,
\begin{equation}
 {1\over 12} S_8 = K^S - \widetilde{K^S}\; ,
 \qquad\qquad 
 {1\over 12} S_{19} = K^S - \overline{K^S}\; ,
 \qquad\qquad 
 {1\over 12} S_{21} = J^S - \overline{J^S}\; ,
\end{equation}
\begin{equation}
 2 S_{11} + S_{10} + S_5 =  {{29}\over 8} - {{35\,{{{ Y_1}^3}}}\over {864}} + 
   {{{ Y_1}}^2}\,\left( {{65}\over {96}} + 
      {{5\,{ Y_{2,1}}}\over {41472}} \right)  + {{{ Y_{2,1}}}\over {576}} + 
   {{{ \left(Y_{2,1}\right)^2}}\over {165888}} - 
   { Y_1}\,\left( {{83}\over {32}} + {{11\,{ Y_{2,1}}}\over {6912}} +
      {{{\left(Y_{2,1}\right)^2}}\over {1990656}} \right) \;,
\label{relationship1}
\end{equation}
\begin{eqnarray}
&& S_1 - S_{10} - 4 S_3 - 2 S_7 + {2\over 3} S_4 = \nonumber \\
  && \quad -{{67}\over {32}} + {{41\,{{{ Y_1}}^3}}\over {432}} + 
   {{{ Y_1}}^2}\,\left( -{{169}\over {192}} + 
      {{7\,{ Y_{2,1}}}\over {82944}} \right)  - {{7\,{ Y_{2,1}}}\over {4608}} - 
   {{{\left( Y_{2,1}\right)^2}}\over {663552}} + 
   {{{\left( Y_{2,1}\right)^3}}\over {286654464}} + 
   { Y_1}\,\left( {{39}\over {16}} + {{{ Y_{2,1}}}\over {1728}} - 
      {{{\left( Y_{2,1}\right)^2}}\over {995328}} \right) \;.
\end{eqnarray}
We give the proof of the relationship Eq.(\ref{relationship1}). 
Twice the numerator in
$S_{11}$ plus the numerator in $S_{10}$ is equal to 
\begin{equation}
 \Delta_{p,q}\,\Delta_{k,r}\left(\Pi_{p+k}+\Pi_{p+q}+\Pi_{p+r}\right)\;,
\end{equation}
which, by using Eq.(\ref{identities3}) under the integration, becomes
\begin{equation}
 \Delta_{p,q}\,\Delta_{k,r}\left(4\,\Pi_p - \Sigma^S_{pqkr}\right)\;.
\end{equation} 
Therefore the following expression is true
\begin{equation}
 2 S_{11} + S_{10} = 4 \int D_3 \;\;{\Delta_{p,q}\,\Delta_{k,r}\over
         \Pi_q\;\Pi_k\;\Pi_r} - S_5 \;.
\end{equation}
The numerator in the integrand can be rewritten after some 
straightforward algebra under the integration
\begin{equation}
 \Delta_{p,q}\,\Delta_{k,r} = -{1\over 2} \left(\widehat{k}^2\right)^2\,
               \widehat{r}^2 +{1\over 4} \widehat{q}^2 \,
               \left(\widehat{k+r}^2\right)^2 + {1\over 8}
               \left(\widehat{q}^2\right)^2\,\widehat{k}^2\,\widehat{r}^2\;,
\end{equation}
and now a use of the second identity in Eq.(\ref{identities2}) yields
the final result shown in Eq.(\ref{relationship1}).

\section{Appendix~C}
{\centerline {\bf Methods for the numerical calculation of the integrals}}.
\vskip 5mm

We have used three methods to calculate the various finite integrals
listed in Appendix~A. They are: {\it i)} an extrapolation to infinite
size of the results obtained at small sizes, {\it ii)} the Gauss
integration and {\it iii)} an extension of the coordinate 
space method~\cite{coord} with the Symanzik propagators. 
Several integrals were evaluated by using more than one method
for checking purposes.
In this Appendix we will briefly describe methods {\it i)} and {\it iii)}. 

In the first method we calculated the integral $I$ at small lattices $L$
obtaining $I(L)$ for several $L$. 
Then we extrapolated this set of results to infinite
size. The extrapolating formula was
\begin{equation}
I(L)\;=\;I(L=\infty)\;+\;{b_1\over L^m} + {b_2\log L\over L^m} \;,
\label{fitf}
\end{equation}
where the correct result for the integral is $I(L=\infty)$. The
exponent in the denominators is 
$m=1$ for the one--loop integrals and $m=2$ for all other integrals.
The errors were determined by looking at the stability of the 
figures against the increase of $L$. In fact such a stability indicates
that further terms in the expansion~Eq.~(\ref{fitf}) are irrelevant for
the last stable digit. For one--loop integrals we obtained
actually many significant digits by working up to lattice sizes as
large as $L=10000$. For the $S_i$ integrals the largest size was 
$L=60$. 

 For finite $L$ the integral $I(L)$ is actually a sum of terms. In
this sum we have excluded the momenta which lead to vanishing
propagators $(\widehat{p}^2=0$ or $\Pi_p=0)$. In the limit
$L\rightarrow \infty$ this procedure implies the exclusion of a region
that has zero measure in the corresponding integration and therefore
it has no consequences on the final extrapolated result
$I(L=\infty)$. We have checked this statement by repeating the
calculation on several one loop integrals firstly {\it i)} by
excluding only the zero mode $\widehat{p}^2=0$ and secondly {\it ii)}
by excluding the lines $\widehat{p}^2_1=0$ or
$\widehat{p}^2_2=0$. Both methods yielded exactly the same final
extrapolated number.

The integrals $H_i$ and $H^S_i$ emerge in the calculation of the
diagram 10 of Fig.~\ref{fig:3}. They are four--loop integrals and as a
consequence a direct application of the above method is rather slow
and a poor precision is obtained. In order to simplify the evaluation
of these four--loop integrals, one can take advantage of the
topology of the diagram and rewrite them as an effective two--loop integral
\begin{equation}
 H_i = \int_{-\pi}^{+\pi} {{{\rm d^2} q} \over  \left(2 \pi\right)^2}
  \sum_k \left( \prod_{j=1}^3 
  \int_{-\pi}^{+\pi} {{{\rm d^2} p} \over  \left(2 \pi\right)^2}
  {{\cal N}_i^{kj}(p,q-p) \over \widehat{p}^2 \;\widehat{(q-p)}^2}\right)\;,
\label{furbo}
\end{equation}
and analogously for $H^S_i$. The sum $\sum_k$ over numerators 
${\cal N}_i^{kj}(p,q-p)$ contains one term in 
$H_5$ and 64 terms in $H^S_3$. Nevertheless, in all cases the integration of 
Eq.(\ref{furbo}) is much faster than a direct integration of 
Eqs.(\ref{h1})--(\ref{hs5}). The largest lattice size used for the
evaluation of these integrals with the method of Eq.(\ref{furbo})
was $L=400$.

In some cases we have also applied the coordinate space method to
check our results.
We need to extend this method, introduced in~\cite{coord,shin},
to include the case of improved propagators. 
Here we will show with some detail the 
calculation of $J^S$. This requires the previous
evaluation of the improved free propagator $G(x)$ 
\begin{equation}
 G(x) \equiv \int_{-\pi}^{+\pi} 
             {{{\rm d^2} p} \over  \left(2 \pi\right)^2}
 {{\rm e}^{i\,p\,x} - 1 \over \Pi_p} \, .
\label{gcoord}
\end{equation}
In terms of $G(x)$ the integral $J^S$ reads
\begin{equation}
 J^S = {1 \over 9} \sum_x \left(16\;
       \sum_{\mu\nu} \left(\partial^+_\mu \partial^+_\nu G(x)\right)^4
       - 2 \sum_{\mu\nu} \left(\partial^+_\nu \left(
       \partial^+_\mu + \partial^-_\mu\right)G(x)\right)^4 +
       {1\over 16} \sum_{\mu\nu} \left(
       \left(\partial^+_\nu + \partial^-_\nu\right)
       \left(\partial^+_\mu + \partial^-_\mu\right) G(x) \right)^4 \right)\,,
\label{jscoord}
\end{equation}
where $\partial^\pm$ have been defined in Eq.(\ref{operators}). 
The procedure is analogous to what is done for the standard case, the only
new features arise because of the presence of an improved Symanzik
propagator. This propagator, Eq.(\ref{gcoord}), satisfies
\begin{equation}
 {\cal L} \,G(x) = - \delta^2(x) \, ,
\label{equation}
\end{equation}
where $\delta^2(x)$ is the Kronecker function on a two--dimensional
lattice and ${\cal L}$ is the improved Laplacian
\begin{equation}
 {\cal L}\equiv {4\over 3} \; \sum_\mu\; \partial^+_\mu \partial^-_\mu -
                {1\over 12}\;\sum_\mu\; \left( \partial^+_\mu + 
                \partial^-_\mu \right)^2 \, .
\label{laplacian}
\end{equation}
If $x\not= 0$ Eq.(\ref{equation}) provides
\begin{equation}
 \sum_\mu G(x+2\widehat\mu) = - \sum_\mu G(x-2\widehat\mu) + 
  4 G(x) + 16 \sum_\mu \left[ G(x+\widehat\mu) + G(x-\widehat\mu) -
  2 G(x) \right]\, .
\label{recurrence}
\end{equation}
Now, by using 
\begin{equation}
 {\partial \over \partial \,p_\mu} \log \Pi_p =
       {1\over \Pi_p} \left( {8\over 3}\, \sin p_\mu - {1\over 3} 
       \,\sin 2 p_\mu\right)\, ,
\end{equation}
we obtain
\begin{equation}
 {4\over 3} \left( G(x+\widehat\mu) - G(x-\widehat\mu)\right) -
 {1\over 6} \left( G(x+2\widehat\mu) - G(x-2\widehat\mu) \right) =
 x_\mu \, X^S(x) \,,
\label{recurrence2}
\end{equation}
where
\begin{equation}
 X^S(x) \equiv
 \int_{-\pi}^{+\pi} {{{\rm d^2} p} \over  \left(2 \pi\right)^2}
 {\rm e}^{i\,p\,x} \log \Pi_p \;.
\end{equation}
Eq.(\ref{recurrence2}) provides a recurrence relation for the
propagator,
\begin{equation}
 G(x+2\widehat\mu)=G(x-2\widehat\mu)+ 8\;\left[
 G(x+\widehat\mu)- G(x-\widehat\mu)\right] - 6 x_\mu X^S(x)
\label{r1}
\end{equation}
which must be complemented with the relationship
\begin{equation}
  X^S(x)={1\over \sum_\mu x_\mu} 
  \left( 10\; G(x) + \sum_\mu \left( {1\over 3} \,
  G(x-2\widehat\mu) -4\; G(x-\widehat\mu) - {4\over 3}\, G(x+\widehat\mu)
  \right)\right)\, ,
\label{r2}
\end{equation}
obtained from Eq.(\ref{recurrence}) and (\ref{recurrence2})
after summing over $\mu$.

\begin{figure}[htbp]
\centerline{\epsfig{file=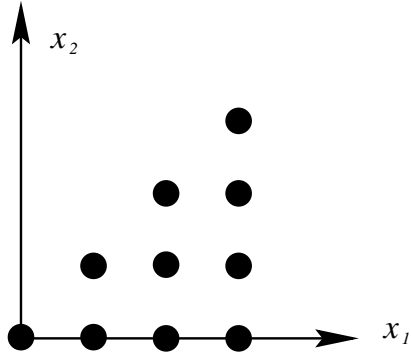,width=0.3\textwidth}}
\caption{The basic lattice sites for the evaluation of the propagator
in the coordinate space method.}
\label{fig:7}
\end{figure}

 From the recurrence Eqs.(\ref{r1})--(\ref{r2}) and the symmetry
relations $G(x_1,x_2)=G(x_2,x_1)$, $G(x_1,x_2)=G(-x_1,x_2)$ ($x_1$ and
$x_2$ are the two components of the coordinate site $x$) we can
determine $G(x)$ for any $x$ in terms of the values of $G(x)$ at 
a basic set of 10 sites. These sites are 
shown in Fig.~\ref{fig:7}. Below we give the values of $G(x)$ on these
points in terms of $Y_{i,j}$ and  $Y_i$
\begin{eqnarray}
 G(0,0)&=& 0 
 \, , \nonumber \\
 G(1,0)&=& {1\over 48} Y_1 - {1\over 4}
 \, , \nonumber \\
 G(1,1)&=& {1\over 1152} Y_{2,1} - {1\over 12} Y_1 - {1\over 8}
 \, , \nonumber \\
 G(2,0)&=& {1\over 3} Y_1 - 1
 \, , \nonumber \\
 G(2,1)&=& {1\over 27648} Y_{3,1} - {1\over 576} Y_{2,1} - {7\over 48}
 Y_1 - {5\over 576}
 \, , \nonumber \\
 G(2,2)&=& {1\over 331776} Y_{4,1} - {1\over 1728} Y_{3,1} +
 {5\over 144} Y_{2,1} - {4\over 3} Y_1 + {917\over 576}
 \, , \nonumber \\
 G(3,0)&=& {-1\over 13824} Y_{3,1} + {1\over 32} Y_{2,1} +
 {27\over 16} Y_1 - {1363\over 288}
 \, , \nonumber \\
 G(3,1)&=& {1\over 1728} Y_{3,1} - {7\over 128} Y_{2,1} + 
 {7\over 12} Y_1 - {19\over 72}
 \, , \nonumber \\
 G(3,2)&=& {1\over 7962624} Y_{5,1} - {1\over 55296} Y_{4,1} +
 {7\over 27648} Y_{3,1} + {1\over 64} Y_{2,1} - {119\over 48} 
 Y_1 + {233341 \over 55296}
 \, , \nonumber \\
 G(3,3)&=& {1\over 95551488} Y_{6,1} - {1\over 331776} Y_{5,1}
 + {1\over 3072} Y_{4,1} - {17\over 864} Y_{3,1} + {89\over 128}
 Y_{2,1} - {27\over 4} Y_1 + {169681\over 497664}\, .
\label{10g}
\end{eqnarray}

\vskip 5mm

We must know the values of $G$ on the above set of sites very
accurately but the above results for $Y_{i,j}$ are not precise
enough (see Table~1). This goal is achieved by using the recurrence
Eqs.(\ref{r1})--(\ref{r2}) backwards. 

 In some cases, we also need the calculation of the squared propagator
$G_2(x)$ defined as
\begin{equation}
 G_2(x)\equiv \int_{-\pi}^{+\pi} 
             {{{\rm d^2} p} \over  \left(2 \pi\right)^2} \;
 {{\rm e}^{i\,p\,x} - 1 +{1\over 2} \sum_\mu \Pi_p^\mu x_\mu^2 
        \over \left(\Pi_p\right)^2} \,.
\label{g2coord}
\end{equation}
The quantity $\Pi_p^\mu$ has been defined in
Eq.(\ref{propagatorsym}). The laplacian Eq.(\ref{laplacian}) applied
on $G_2$ produces ${\cal L}G_2(x) = - G(x)$, or alternatively
\begin{equation}
 \sum_\mu G_2(x+2\widehat\mu) = - \sum_\mu G_2(x-2\widehat\mu) + 
  4 G_2(x) + 16 \sum_\mu \left[ G_2(x+\widehat\mu) + G_2(x-\widehat\mu) -
  2 G_2(x) \right] + 12\;G(x)\, .
\label{recurg2}
\end{equation} 
Now, by using that
\begin{equation}
 {\partial \over \partial p_\mu} {1\over \Pi_p} =
 - {1\over \left(\Pi_p\right)^2} \left( {8\over 3}\, \sin p_\mu - {1\over 3} 
       \,\sin 2 p_\mu\right)\, ,
\end{equation}
we obtain the basic recurrence relation
\begin{equation}
 G_2(x+2\widehat\mu)=G_2(x-2\widehat\mu)+ 8\;\left[
 G_2(x+\widehat\mu)- G_2(x-\widehat\mu)\right] + 6 x_\mu X_2^S(x)\;,
\label{recur2g2}
\end{equation}
where $X_2^S(x)= G(x) + 1/4\pi$. Summing this equation over
$\mu$ and using Eq.(\ref{recurg2}), we get 
\begin{equation}
 X_2^S(x) = {1\over \sum_\mu x_\mu} \left(2 \; G(x) 
 -10\; G_2(x) - {1\over 3} \; \sum_\mu
 G_2(x-2\widehat\mu) + {4\over 3}\;\sum_\mu G_2(x+\widehat\mu) + 4\;
 \sum_\mu G_2(x-\widehat\mu)\right) \; ,
\end{equation}
which must be complemented with Eq.(\ref{recur2g2}) to obtain $G_2(x)$
for any $x$ from the same basic set of points shown in
Fig.~\ref{fig:7}. 

Further tricks necessary for the computation of the integrals are 
analogous to those already explained in Ref.~\cite{coord}.

\section{Appendix~D}
{\centerline {\bf Form of $w^{\rm 1-Symanzik}_4$}}
\vskip 5mm

The expression for $w^{\rm 1-Symanzik}_4$ can be written as the sum of
$w^{\rm 0-Symanzik}_4$ plus a pure 1--loop Symanzik action contribution.
This contribution contains terms proportional to $(n-1)$, $(n-1)^2$
and several powers of the coefficients $c_i$, $i=5,...\;,9$. Let us parametrize
these contributions in the following way
\begin{eqnarray}
 w^{\rm 1-Symanzik}_4 = w^{\rm 0-Symanzik}_4 &+&
  {3\over 2}(n-1)^2 \;\Big( c_5 \;q_5 + 
  c_6 \;q_6+ c_7 \;q_7+ c_8 \;q_8+ c_9 \;q_9+ 
                 c_5\;c_7 \;q_{57}+ c_5\;c_8 \;q_{58} 
  \nonumber \\
 && \qquad\quad \qquad
  + c_5\;c_9 \;q_{59} + c_6\;c_7 \;q_{67}+ c_6\;c_8 \;q_{68} + 
         c_6\;c_9 \;q_{69}\Big) 
  \nonumber \\
  &+& {3\over 2}(n-1) \;\Big( c_5 \;p_5+ c_5^2 \;p_{55} + c_5^3\;p_{555} + 
         c_6 \;p_6 + c_6^2 \;p_{66}+ c_6^3 \;p_{666}
  \nonumber \\
  && \qquad\quad\qquad
        + c_5\;c_6 \;p_{56}+ c_5^2\;c_6 \;p_{556} + c_5\;c_6^2 \;p_{566} + 
         c_7 \;p_7+ c_8 \;p_8+ c_9 \;p_9
  \nonumber \\
  && \qquad\quad\qquad
        + c_5\;c_7 \;p_{57}+ c_5\;c_8 \;p_{58}+ c_5\;c_9 \;p_{59}+ 
         c_6\;c_7 \;p_{67}+ c_6\;c_8 \;p_{68}+ c_6\;c_9 \;p_{69}\Big)\;.
\end{eqnarray}
Now the several coefficients $q_i$ are
\begin{eqnarray}
 q_5&=& {1\over 2} S_{18} - {1\over 4} S_1 - S_2 - {1\over 2} S_3
        \;,\nonumber \\
 q_6 &=& {1\over 2} S_{19} - {1\over 4} S_8 
        \;,\nonumber \\
 q_7 &=& -{1\over 4} - {1\over 4} S_{10} + {19\over 48} Y_1 - {67\over 576}
         {Y_1}^2 + {49\over 6912} {Y_1}^3 - {1\over 72} Y_2 
         + {1\over 144} Y_1 \,Y_2 - {5\over 10368} {Y_1}^2\,Y_2
        \;,\nonumber \\
 q_8 &=& -{1\over 8} - {1\over 4} S_{12} + {19\over 96} Y_1 - {67\over 1152}
         {Y_1}^2 + {49\over 13824} {Y_1}^3 - {1\over 144} Y_2 
         + {1\over 288} Y_1 \,Y_2 - {5\over 20736} {Y_1}^2\,Y_2
        \;,\nonumber \\
 q_9 &=& -{1\over 2} - {1\over 16} \left( S_{11} + S_{14} - 2 S_{15}\right) + 
         {2\over 3} Y_1 - {41\over 144}
         {Y_1}^2 + {17\over 432} {Y_1}^3 - {1\over 36} Y_2 
         + {1\over 48} Y_1 \,Y_2 - {5\over 1296} {Y_1}^2\,Y_2
        \;,\nonumber \\
 q_{57} &=& 8 - {2\over 3} Y_1 + {1\over 12} Y_{2,1} - {1\over 144} 
            Y_1 \, Y_{2,1} - {1\over 432} Y_{3,2} + {1\over 5184}
            Y_1\, Y_{3,2}
        \;,\nonumber \\
 q_{58} &=& 4 - {1\over 3} Y_1 + {1\over 24} Y_{2,1} - {1\over 288} 
            Y_1 \, Y_{2,1} - {1\over 864} Y_{3,2} + {1\over 10368}
            Y_1\, Y_{3,2}
        \;,\nonumber \\
 q_{59} &=& -2 + {2\over 3} Y_1 + {1\over 8} Y_{2,1} - {1\over 24} 
            Y_1 \, Y_{2,1} - {1\over 216} Y_{3,2} + {1\over 648}
            Y_1\, Y_{3,2}
        \;,\nonumber \\
 q_{67} &=& 4 Y_1 - {1\over 3} {Y_1}^2 - {1\over 3} Y_2 + {1\over 36} 
           Y_1\,Y_2
        \;,\nonumber \\
 q_{68} &=& 2 Y_1 - {1\over 6} {Y_1}^2 - {1\over 6} Y_2 + {1\over 72} 
           Y_1\,Y_2
        \;,\nonumber \\
 q_{69} &=& 2 Y_1 - {2\over 3} {Y_1}^2 - {2\over 3} Y_2 + {2\over 9} 
           Y_1\,Y_2\;,
\end{eqnarray}
and those $p_i$ are
\begin{eqnarray}
 p_5 &=& - {123\over 128}  + {1\over 6} S_{20} - {1\over 2} S_{18}
         + {1\over 4} S_1 + S_2 + {1\over 2} S_3 - {1\over 3} S_4 
         + {1\over 4} S_5 + S_6 + S_7 +{93 \over 64} Y_1 -
         {185 \over 288} {Y_1}^2 + {215 \over 2304} {Y_1}^3 
        \nonumber \\
     && + {353\over 18432} Y_{2,1} - {121\over 6912} Y_1\,Y_{2,1}
        +{655 \over 165888} {Y_1}^2 \,Y_{2,1} - {11\over 884736}
         \left(Y_{2,1}\right)^2 + {7\over 1327104} Y_1\,\left(Y_{2,1}\right)^2 +
         {1\over 1146617856} \left(Y_{2,1}\right)^3
        \nonumber \\
     && - {5\over 48} Y_2 + {145 \over 1728}Y_1\,Y_2 - {175\over 10368}
        {Y_1}^2 \,Y_2 + {5\over 10368}Y_2\,Y_{2,1} -
         {25\over 124416} Y_1\,Y_2\,Y_{2,1} - {11\over 6912}Y_{3,2}
        \nonumber \\
     && +{77\over 55296} Y_1 \, Y_{3,2}-{605\over 1990656} {Y_1}^2\,
        Y_{3,2} + {1\over 1990656}Y_{2,1} \,Y_{3,2} - {5\over 23887872}
        Y_1\,Y_{2,1}\,Y_{3,2}
        \nonumber \\
     && -{5\over 248832} Y_2\,Y_{3,2} + {25\over 2985984}Y_1\,Y_2\,Y_{3,2}
        +{1\over 20736}Y_{4,3} - {5\over 124416}Y_1\,Y_{4,3} +
        {25\over 2985984} {Y_1}^2 \,Y_{4,3}\;,
       \nonumber \\
  p_{55} &=& -{557\over 48} + {425\over 72} Y_1 + {349\over 768}
             Y_{2,1} - {25\over 144} Y_1\,Y_{2,1} - {7\over 6912}
             \left(Y_{2,1}\right)^2 + 
             {13\over 1728} Y_{3,1} -{5\over 1728}Y_1\,Y_{3,1}
       \nonumber \\
     && -{1\over 248832} Y_{2,1}\,Y_{3,1} -{5\over 288}Y_{3,2} +
         {35\over 5184} Y_1\,Y_{3,2} + {5\over 62208}Y_{2,1} Y_{3,2}-
         {5\over 2985984}\left(Y_{3,2}\right)^2 -{11\over 27648}Y_{4,2}
       \nonumber \\
     && +{5\over 31104} Y_1\,Y_{4,2}+{1\over 11943936}Y_{2,1}\,Y_{4,2}
        +{1\over 124416} Y_{5,3} - {5\over 1492992}Y_1\,Y_{5,3} \;,
       \nonumber \\
  p_{555}&=& {1670\over 27} + {5\over 5184} Y_{4,1} - {1\over 31104}
           Y_{5,2} + {1\over 2239488} Y_{6,3} \;,
       \nonumber \\
  p_6&=& -{5\over 8} +{1\over 6} S_{21} - J^S +\widetilde{J^S} -
        {1\over 2} S_{19} + {1\over 4} S_8 + {41\over 32} Y_1 
         -{95\over 128}{Y_1}^2 +{1889\over 13824} {Y_1}^3 -{5\over 32} Y_2
       \nonumber \\
     && +{29\over 192} Y_1\,Y_2 - {55\over 1536} Y_2 {Y_1}^2 -
        {5\over 1728} {Y_2}^2 +{25\over 20736} Y_1\,{Y_2}^2 +
        {1\over 144} Y_3 - {5\over 864} Y_1\,Y_3 +{25\over 20736}
        {Y_1}^2 Y_3 \;,
       \nonumber \\
 p_{66}&=& 6Y_1 -{5\over 2} \left({Y_1}^2 +Y_2\right) +{29\over 24}
           Y_1\,Y_2 - {5\over 144}{Y_2}^2 + {1\over 6} Y_3 -
           {5\over 72} Y_1 \,Y_3 \;,
       \nonumber \\
 p_{666} &=& {4\over 3} Y_3 \;,
       \nonumber \\
 p_{56} &=& -15 + 12 Y_1 - {5\over 3} {Y_1}^2 +{41\over 48}Y_{2,1} 
            -{35\over 96} Y_1\,Y_{2,1} -{1\over 3456} \left(Y_{2,1}\right)^2
            -{5\over 2} Y_2 +{35\over 36}Y_1\,Y_2 +{5\over 432}
            Y_2\,Y_{2,1} 
       \nonumber \\
      && - {43\over 576} Y_{3,2} +{109\over 3456}Y_1\,Y_{3,2} 
          +{1\over 82944} Y_{2,1}\,Y_{3,2} -{5\over 10368}
          Y_2\,Y_{3,2} +{1\over 432}Y_{4,3} - {5\over 5184} Y_1\,Y_{4,3}\;,
       \nonumber \\
 p_{556} &=& 44 + {1\over 6} Y_{3,1} - {1\over 108} Y_{4,2} +
            {1\over 5184} Y_{5,3} \;,
       \nonumber \\
 p_{566} &=& 4 Y_{2,1} - {2\over 3} Y_{3,2} + {1\over 36} Y_{4,3}\;,
       \nonumber \\
 p_7 &=& {1\over 2} - {1\over 2} S_{11} + {1\over 12} Y_1 -
         {275\over 1152} {Y_1}^2 + {215\over 3456} {Y_1}^3
         -{1\over 192} Y_{2,1} + {19\over 4608} Y_1\,Y_{2,1}
       \nonumber \\
    &&   -{145\over 165888}{Y_1}^2 \,Y_{2,1} -{1\over 82944}
         \left(Y_{2,1}\right)^2
        +{11\over 1990656}Y_1\,\left(Y_{2,1}\right)^2 -{25\over 288} Y_2 +
       {29\over 384} Y_1\,Y_2 -{85\over 5184} {Y_1}^2 Y_2 
       \nonumber \\
    && -{1\over 13824}Y_{2,1}\,Y_2 +{5\over 165888} Y_1\,Y_2\,Y_{2,1} 
       +{1\over 4608} Y_{3,2} -{1\over 6144} Y_1\,Y_{3,2} +
       {5\over 165888} {Y_1}^2 \,Y_{3,2} 
       \nonumber \\
    && + {1\over 1990656} Y_{2,1} \,Y_{3,2} -
        {5\over 23887872} Y_1\,Y_{2,1}\,Y_{3,2}\;,
       \nonumber \\
 p_8 &=&  {1\over 4} - {1\over 2} S_{13}+{7\over 48} Y_1 - {121\over 576}
          {Y_1}^2+{353 \over 6912} {Y_1}^3 - {1\over 18} Y_2 +
          {5\over 96} Y_1\,Y_2 - {125\over 10368} {Y_1}^2 \,Y_2\;,
       \nonumber \\
 p_9&=& -{9\over 8} - {1\over 16}\left(S_{10} +S_{11} +S_{14} - 2 S_{15} +
        S_{16} - 2 S_{17} \right) +{3\over 2} Y_1 - {31\over 48} {Y_1}^2 +
        {13\over 144} {Y_1}^3 + {1\over 1152} Y_{2,1} 
       \nonumber \\
    && -{1\over 1728} Y_1\,Y_{2,1} +{1\over 10368} {Y_1}^2 \,Y_{2,1}
      -{1\over 12} Y_2 + {1\over 16} Y_1\,Y_2 - {5\over 432} 
        {Y_1}^2 \,Y_2\;,
      \nonumber \\
 p_{57} &=& -{43\over 16} +{187\over 48} Y_1 + {871\over 2304} 
          Y_{2,1} - {1\over 6} Y_1\,Y_{2,1} +{1\over 3456} 
          \left(Y_{2,1}\right)^2
          -{1\over 576} Y_{3,1} +{1\over 1728} Y_1\,Y_{3,1} 
      \nonumber \\
   && -{1\over 248832} Y_{2,1}\,Y_{3,1} -{25\over 1728} Y_{3,2} +
      {17\over 2592} Y_1\,Y_{3,2} -{1\over 82944} Y_{2,1} \,
      Y_{3,2} +{1\over 27648} Y_{4,2} 
      \nonumber \\
   && -{1\over 82944} Y_1\,Y_{4,2} + {1\over 11943936} Y_{2,1} \,Y_{4,2}\;,        
      \nonumber \\
 p_{58} &=& -4 +{13\over 3} Y_1 + {1\over 4} Y_{2,1} - {1\over 8} 
         Y_1\,Y_{2,1} -{1\over 108} Y_{3,2} +{25\over 5184} 
         Y_1\,Y_{3,2} \;,
      \nonumber \\
 p_{59} &=& -6 + 2Y_1 +{3\over 8} Y_{2,1} -{1\over 8} Y_1\,Y_{2,1} 
           -{1\over 72} Y_{3,2} + {1\over 216} Y_1\,Y_{3,2} \;,
      \nonumber \\
 p_{67} &=& 9 +Y_1 - {4\over 3} {Y_1}^2 -{5\over 48} Y_{2,1} +
        {1\over 24} Y_1\,Y_{2,1} - {1\over 3456} \left(Y_{2,1}\right)^2 -
        {25\over 12} Y_2 +{17\over 18} Y_1\,Y_2 
      \nonumber \\
  &&  - {1\over 576} Y_2 \,Y_{2,1} + {1\over 192} Y_{3,2} -
      {1\over 576} Y_1\,Y_{3,2} + {1\over 82944} Y_{2,1}\,Y_{3,2}\;,
      \nonumber \\
 p_{68} &=& 4 Y_1 - {4\over 3} {Y_1}^2 - {4\over 3}Y_2 +
        {25\over 36} Y_1\,Y_2 \;,
      \nonumber \\
 p_{69} &=& 6 Y_1 - 2 {Y_1}^2 - 2 Y_2 +
        {2\over 3} Y_1\,Y_2 \;.
\end{eqnarray}
Here a new set of integrals has appeared, $S_1$, $S_2$, etc.,
coming from diagrams 4, 5 of Fig.~\ref{fig:6}. They are defined 
and calculated in Appendix~A. This set of integrals is not completely
independent and in Appendix~B we give some relationships among them.

\end{document}